\documentstyle[psfig]{mn}

\newif\ifAMStwofonts
\ifoldfss
  \ifCUPmtlplainloaded \else
    \NewTextAlphabet{textbfit} {cmbxti10} {}
    \NewTextAlphabet{textbfss} {cmssbx10} {}
    \NewMathAlphabet{mathbfit} {cmbxti10} {} % for math mode
    \NewMathAlphabet{mathbfss} {cmssbx10} {} %  "   "    "
  \fi
  \ifAMStwofonts
    \ifCUPmtlplainloaded \else
      \NewSymbolFont{upmath} {eurm10}
      \NewSymbolFont{AMSa} {msam10}
      \NewMathSymbol{\upi}     {0}{upmath}{19}
      \NewMathSymbol{\umu}     {0}{upmath}{16}
      \NewMathSymbol{\upartial}{0}{upmath}{40}
      \NewMathSymbol{\leqslant}{3}{AMSa}{36}
      \NewMathSymbol{\geqslant}{3}{AMSa}{3E}

       \let\le=\leqslant
       \let\ge=\geqslant
    \fi
  \fi
\fi % End of OFSS

\ifnfssone
  \newmathalphabet{\mathit}
  \addtoversion{normal}{\mathit}{cmr}{m}{it}
  \addtoversion{bold}{\mathit}{cmr}{bx}{it}
  \newmathalphabet{\mathbfit} % math mode version of \textbfit{..}
  \addtoversion{normal}{\mathbfit}{cmr}{bx}{it}
  \addtoversion{bold}{\mathbfit}{cmr}{bx}{it}
  \newmathalphabet{\mathbfss} % math mode version of \textbfss{..}
  \addtoversion{normal}{\mathbfss}{cmss}{bx}{n}
  \addtoversion{bold}{\mathbfss}{cmss}{bx}{n}
  \ifAMStwofonts
    \ifCUPmtlplainloaded \else
      \UseAMStwoboldmath
      \makeatletter
      \new@mathgroup\upmath@group
      \define@mathgroup\mv@normal\upmath@group{eur}{m}{n}
      \define@mathgroup\mv@bold\upmath@group{eur}{b}{n}
      \edef\UPM{\hexnumber\upmath@group}
      \new@mathgroup\amsa@group
      \define@mathgroup\mv@normal\amsa@group{msa}{m}{n}
      \define@mathgroup\mv@bold\amsa@group{msa}{m}{n}
      \edef\AMSa{\hexnumber\amsa@group}
      \makeatother
      \mathchardef\upi="0\UPM19
      \mathchardef\umu="0\UPM16
      \mathchardef\upartial="0\UPM40
      \mathchardef\leqslant="3\AMSa36
      \mathchardef\geqslant="3\AMSa3E

       \let\le=\leqslant
       \let\ge=\geqslant
    \fi
  \fi
\fi % End of NFSS release 1

\ifnfsstwo
  \DeclareMathAlphabet{\mathbfit}{OT1}{cmr}{bx}{it}
  \SetMathAlphabet\mathbfit{bold}{OT1}{cmr}{bx}{it}
  \DeclareMathAlphabet{\mathbfss}{OT1}{cmss}{bx}{n}
  \SetMathAlphabet\mathbfss{bold}{OT1}{cmss}{bx}{n}
  \ifAMStwofonts
    \ifCUPmtlplainloaded \else
      \DeclareSymbolFont{UPM}{U}{eur}{m}{n}
      \SetSymbolFont{UPM}{bold}{U}{eur}{b}{n}
      \DeclareSymbolFont{AMSa}{U}{msa}{m}{n}
      \DeclareMathSymbol{\upi}{0}{UPM}{"19}
      \DeclareMathSymbol{\umu}{0}{UPM}{"16}
      \DeclareMathSymbol{\upartial}{0}{UPM}{"40}
      \DeclareMathSymbol{\leqslant}{3}{AMSa}{"36}
      \DeclareMathSymbol{\geqslant}{3}{AMSa}{"3E}

       \let\le=\leqslant
       \let\ge=\geqslant
    \fi
  \fi
\fi % End of NFSS release 2

\ifCUPmtlplainloaded \else
  \ifAMStwofonts \else % If no AMS fonts
    \def\upi{\pi}
    \def\umu{\mu}
    \def\upartial{\partial}
  \fi
\fi

\title{Galaxy threshing and the origin of ultra-compact dwarf galaxies in the Fornax cluster}
\author[K. Bekki,  W.  J. Couch, M. J. Drinkwater, Y. Shioya]
       {K. Bekki,${}^1$, 
        W. J. Couch${}^1$
        M. J. Drinkwater${}^2$, 
        and Y. Shioya${}^3$\\
        ${}^1$School of Physics, University of New South Wales, Sydney 2052, NSW, Australia \\
        ${}^2$Department of Physics, University of Queensland, Queensland 4072, Australia\\
        ${}^3$Astronomical Institute, Tohoku University, Sendai, 980-8578, Japan} 
\date{Accepted 
      Received
      in original form 2001}

\pagerange{\pageref{firstpage}--\pageref{lastpage}}
\pubyear{1994}

\begin{document}

\maketitle

\label{firstpage}

\begin{abstract}

A recent all-object spectroscopic survey
centred on the Fornax cluster of galaxies, has discovered 
a population of sub-luminous and extremely compact members,
called ``ultra-compact dwarf'' (UCD) galaxies.
In order to clarify the origin of these objects, we have used 
self-consistent numerical simulations to study the dynamical evolution
a nucleated dwarf galaxy would undergo if orbiting the center of the 
Fornax cluster and suffering from its strong tidal gravitational field.
We find that the outer stellar components of a nucleated dwarf 
are removed by the strong tidal field of the cluster,  
whereas the nucleus manages to survive as a result of 
its initially compact nature. The developed naked nucleus is found 
to have physical properties (e.g., size and mass) similar to those 
observed for UCDs. We also find that although this formation process does 
not have a strong dependence on the initial total luminosity of 
the nucleated dwarf, it does depend on the radial density profile of 
the dark halo in the sense that UCDs are less likely to be formed from 
dwarfs embedded in dark matter halos with central `cuspy' density 
profiles. Our simulations also suggest that very massive and compact 
stellar systems can be rapidly and efficiently formed in the central 
regions of dwarfs through the merging of smaller star clusters. 
We provide some theoretical predictions on the total number
and radial number density profile of UCDs in a cluster and their  
dependences on cluster masses.
\end{abstract}


\section{Introduction}

The population of low-luminosity, low surface brightness 
dwarf elliptical  and irregular galaxies are thought to have a key
role in the context of galaxy formation and evolution. This is 
not only because in the hierarchical clustering scenario these 
small galaxies merge to form more massive galaxies, but also because 
their observed physical properties place strong constraints on theoretical 
models of galaxy formation and evolution (e.g., Ferguson \& Binggeli 1994; 
Mateo 1998; Grebel 1998). Examples of the latter include 
the scaling relation (Kormendy 1977;  Ferguson \& Binggeli
1994), the luminosity function (Binggeli, Sandage, \& Tammann
1985; Sandage, Binggeli, \& Tammann 1985), the presence of 
nuclear structures (Binggeli \& Cameron 1991),
and rotation-curve profiles (Moore 1994), all of which 
have been extensively discussed for assessing the viability
of any theory of galaxy formation and evolution.

In contrast to this ubiquitous dwarf galaxy population,
the red compact elliptical galaxies (cE) such as M32 
are very rarely observed and are preferentially located in 
close proximity to giant galaxies (e.g., Nieto \& Prugniel 1987).
Only recently, Forbes et al. (2002) have discovered another clear
example of an M32-like compact spheroidal system very close
to an M31-like spiral galaxy, in the image of UGC 10214 taken
by Advanced Camera for Surveys ($ACS$) on board the Hubble Space Telescope 
($HST$). The origin of the high surface brightness and radially-limited 
luminosity profiles of such objects has been discussed by several authors 
mostly in the context of the tidal effects of the massive galaxies they
lie close to (King 1962; Faber 1973; Nieto \& Prugniel 1987; Burkert 1994; 
Bekki et al. 2001a).  It is, however, still controversial how such compact 
density profiles  are formed via the tidal gravitational forces of massive
galaxies. 

A new type of sub-luminous and extremely compact ``dwarf galaxy'' has
recently been discovered in an ``all-object''  spectroscopic survey
centred on the Fornax cluster of galaxies (Drinkwater et al. 2000a, b).
While objects with this type of {\it morphology} have been observed 
before -- the bright compact objects discovered by Hilker et 
al. 1999 -- and the very luminous globular clusters around cD galaxies 
(Harris, Pritchet, \& McClure 1995) -- in this particular case they
have been found to be members of the Fornax cluster, 
have intrinsic sizes of only $\sim$ 100\,pc, and have absolute
$B-$band magnitudes ranging from $-13$ to $-11$\,mag. Hence Drinkwater et 
al. have named them ``ultra-compact dwarf'' (UCD) galaxies.
Importantly, the luminosities of UCDs are intermediate between those of 
globular clusters and small dwarf galaxies and  are similar to those of the 
bright end of the luminosity function of the nuclei of nucleated dwarf 
ellipticals (e.g., Lotz et al. 2001). 
Moreover, they are observed to be within 30\,arcmin of the 
central dominant galaxy in Fornax, NGC~1399, and are distributed as large 
radii as this galaxy's globular cluster system (extending out to
a radius of 100 kpc at least; Dirsch et al. 2003).
Owing to the lack of ultra high-resolution imaging and high-dispersion 
spectroscopy of the UCDs, it was highly uncertain at the time of their
discovery whether they are super-massive star clusters (intracluster 
globular clusters or tidally stripped nuclei of dwarf galaxies) or really a 
new type of low-luminosity compact elliptical dwarf (``M32-type'') galaxy 
(Drinkwater et al. 2000b).

However, Drinkwater et al. (2003; hereafter D03) recently obtained 
high-resolution imaging of the UCDs in Fornax with the Space Telescope 
Imaging Spectrograph (STIS) on $HST$ and found that their radial density 
profiles  (with one exception) can be well fitted by both King model 
and de Vaucouleurs $R^{1/4}$ profiles. The effective radii measured for 
the UCDs were in the $10-30$\,pc range, which is larger than the typical 
core radius of a globular cluster ($\sim$ 6\,pc) and smaller
than the scale length of the most compact normal dwarf galaxy
in the Virgo cluster ($\sim$\,160 pc). D03 also 
obtained high-resolution spectra of UCDs using the Very Large Telescope 
UV Echelle Spectrograph and the Keck Telescope Echelle Spectrograph. These
were used to measure internal velocity dispersions and, in turn, 
mass-to-light ($M/L$) ratios for the UCDs. These ranged from 
24 to 37\,km\,s$^{-1}$ and 2 to 4\,$M/L_{\odot}$, respectively.  
These observations suggest that UCDs are different from normal dwarf galaxies
in size and from GCs in velocity dispersion and $M/L$: UCDs are thus observed
to be really a new class of galactic object!

The importance of the tidal field of more massive galaxies 
in forming globular clusters (and even objects that are an order of magnitude
brighter than globular clusters)  from nucleated dwarf galaxies has
already been discussed by several authors
(e.g., Zinnecker et al. 1988; Freeman 1993; Bassino et al. 1994).
Bassino et al. (1994) first demonstrated that very massive globular clusters 
can originate from nuclei of nucleated dwarf galaxies around more massive
galaxies owing to the tidal stripping of the stellar envelope of the dwarfs.
Using more realistic numerical simulations, Bekki et al. (2001b) proposed 
the ``galaxy threshing  scenario'' in which the global tidal field of the 
Fornax cluster efficiently strips the outer stellar envelopes of the 
more massive nucleated dwarf galaxies (M$_{\rm d}$ $\sim$ $10^8$ M$_{\odot}$) 
as they orbit NGC~1399. This was shown to transform them into 
compact systems whose luminosity was consistent with that of the
UCDs. This previous study, however, did not discuss structural and
kinematical properties (e.g., radial density profile, velocity dispersion,
and mass-to-light-ratio) of which there are now measurements 
from $HST$ and the large ground-based telescopes (D03). 

The purpose of this paper is to numerically investigate the threshing 
scenario in a more comprehensive and self-consistent manner and to
specifically address and compare these structural and kinematical 
properties. Based on a fully self-consistent dynamical model of 
dE,Ns  embedded in massive dark matter halos,
we investigate the following issues:
(i)\,how the global tidal field of the Fornax cluster 
strips the dark matter halos, the stellar envelopes, and the nuclei
of dE,Ns with different luminosites,
(ii)\,whether the structural properties of the dark matter halos of dE'Ns 
are important for the effectiveness of the threshing process, 
(iii)\,whether galaxy threshing can transform nucelated dwarf spirals
into UCDs,
(iv)\,whether the galaxy threshing scenario can explain the relationship 
between velocity dispersion and absolute magnitude observed for UCDs, and
(v)\,what the threshing scenario predicts for the total number and
the radial number distribution of UCDs in a cluster and their dependence
on cluster mass.
We also discuss a possible physical relationship between
UCDs, the globular cluster $\omega$Cen, and the compact elliptical galaxy 
M32 and a formation mechanism for the nuclei of dE,Ns. 

The plan of the paper is as follows: In the next section,
we describe our numerical model for the dynamical evolution
of dE,Ns in the Fornax cluster. 
In \S 3, we present our numerical results on the  
structural and  kinematical properties of UCDs formed
by galaxy threshing and compare these with recent observational results.
In \S 4, we discuss the environmental dependences of UCD properties,
the possible formation of the giant globular cluster $\omega$Cen via 
threshing, and the formation of galactic nuclei in dwarfs via merging of 
globular clusters. We summarise our  conclusions in \S 5.
 
\section{Model Details}

\subsection{Cluster mass profile}

In order to model, numerically, the dynamical evolution of nucleated 
dwarf galaxies as they orbit the centre of the Fornax cluster, we consider 
them to be collisionless stellar systems with a mass and size similar to 
that observed for the dE,N types. We assume that the gravitational field 
of the cluster's dark matter halo has the strongest influence on the 
dynamical evolution of dE,Ns.  Accordingly, we model only the cluster tidal 
field and do not include any tidal effects from other cluster member 
galaxies in the present simulations.
Also cluster potential is assumed to be time-indepedent,
which is not so consistent with recent high-resolution numerical
simulations of cluster galxies formation within the CDM models 
(e.g., Ghigna et al. 1998).
Although our simulations are thus 
idealized in some aspects, we believe that our model still contains the 
essential ingredients that govern the dynamical evolution of dE,Ns in the 
Fornax cluster.

To give our model a realistic radial density profile for
the dark matter halo of the cluster, we base it on both the X-ray 
observations of Jones et al. (1997) and the predictions from the standard cold
dark matter cosmogony (Navarro, Frenk, \& White 1996, hereafter NFW). 
The NFW profile is described as:
\begin{equation}
{\rho}(r)=\frac{\rho_{0}}{(r/r_{\rm s})(1+r/r_{\rm s})^2},
\end{equation}
where $r$,  $\rho_{0}$,  and $r_{\rm s}$ are the distance from the center 
of the cluster, the central density, and the scale-length of the dark halo, 
respectively. The adopted NFW model has a total mass of 
$7.0 \times 10^{13} \rm M_{\odot}$ (within the virial radius) and 
$r_{\rm s}$ of 83\,kpc. This model is based on 
the NFW model with a ``c'' parameter of 12.8.
The adopted set of parameters is consistent with the 
X-ray observations of Jones et al. (1997) and 
the mass estimation of the Fornax cluster derived  from  kinematics of the cluster 
galaxies 
by Drinkwater et al. (2001).
This cluster model is referred to as ``the Fornax profile'' and  
labeled as ``FO''.

To determine the dependences of the galaxy threshing processes on
cluster mass and size, we have also investigated the dynamical evolution
of dE,Ns for other model clusters. In particular, we present the results 
for models with $M_{\rm cl}$ = $5.0 \times 10^{14} \rm M_{\odot}$ and 
$r_{\rm s}$ = 226\,kpc, as well as $M_{\rm cl}$ =  $3.0 \times 10^{13} \rm 
M_{\odot}$ and $r_{\rm s}$ = 83\,kpc. The more massive model 
corresponds to the Virgo cluster, and accordingly it is referred to as 
``VO'' model; the less massive model corresponds to the Local Group and is 
referred to as the ``LG'' model. All of the parameters for these are chosen 
from the values listed in Table of NFW (e.g, model 3 and 15).

\begin{figure*}
\psfig{file=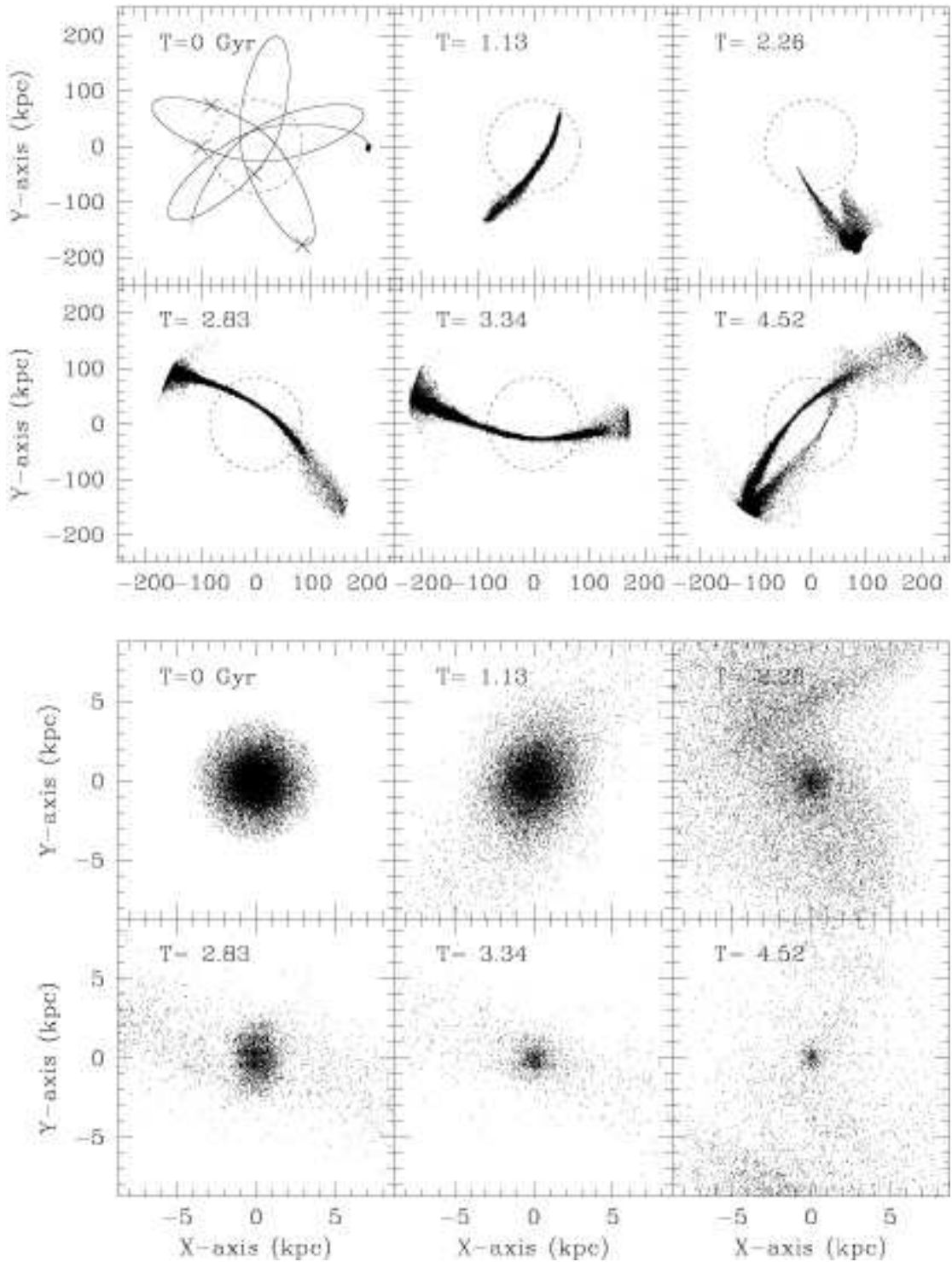}
\caption{
Morphological evolution of the stellar components (the stellar envelope
and the nucleus) of the dE,N projected onto the $x$-$y$ plane for the 
fiducial model (FO1). The time $T$ (in units of Gyr) indicated in the 
upper left corner of each frame represents the time that has elapsed since
the simulation starts. Each frame is 500\,kpc on a side in the upper six  
panels and 17.5\,kpc on a side in the lower panels.
The orbital evolution of the dE,N at 4.5\,Gyr is 
indicated by the solid line in the upper left panel in the 
upper six panels. 
The location of the nucleus of the dwarf is indicated
by crosses for $T$ = 1.13, 2.26, 2.83, and 3.34 Gyr.
The scale radius of the adopted NFW model for the dark matter halo
distribution of the Fornax cluster mass profile
is indicated by a dotted circle in each of the upper six panels.
}
\label{Figure 1}
\end{figure*}

\subsection{Nucleated dwarf ellipticals}

A dE,N is modelled as a fully self-gravitating system
and assumed to consist of a dark matter halo, a stellar component, and
a nucleus. 
We accordingly do not include any gaseous components in
the present models.
For convenience, the stellar component 
(i.e., the main baryonic component) is referred to as either
the ``envelope'' or the ``stellar envelope'' so that 
we can distinguish this component from the stellar nucleus. 
The density profile of the dark matter halo 
of the dE,N  is represented by that proposed by
Salucci \& Burkert (2000):
\begin{equation}
{\rho}_{\rm dm}(r)=\frac{\rho_{dm,0}}{(r+a_{\rm dm})(r^2+a_{\rm dm})^2},
\end{equation} 
where $\rho_{dm,0}$ and $a_{\rm dm}$ are the central dark matter 
density and the core (scale) radius, respectively.
For convenience, we hereafter call this profile the ``SB'' profile (or 
model). The main difference between the SB profile and the NFW profile
is that the former has a large dark matter core.
We choose the SB profile rather than the NFW profile,
because the observed kinematics of dwarf galaxies 
are inconsistent with the cuspy dark matter profiles
predicted by the CDM model (e.g., Moore 1994; Burkert 1995). 
For the SB profile, the dark matter
core parameters, $\rho_{dm,0}$,  $a_{\rm dm}$,  and $M_{0}$
(where $M_{0}$ is the total dark matter mass within $a_{\rm dm}$)
are not free parameters, and clear correlations are observed between
them (Burkert 1995): 
\begin{equation}
M_{0}=4.3 \times 10^7 {(\frac{a_{\rm dm}}{\rm kpc})}^{7/3} M_{\odot}.
\end{equation} 
All dark matter particles are distributed within 5$a_{\rm dm}$.

Equations (2) and (3) predict that $a_{\rm dm}$  is  $\sim$ 3.5 times 
larger than the typical scale length ($\sim$ 800\,pc) observed for dE,Ns,  
with $M_{\rm B}$ = $-16$ mag and  $M/L_{\rm B}$ = 10 (Ferguson \& Binggeli 
1994). This large core radius is a favorable condition for galaxy 
threshing to act on dE,Ns, as is described later.
Since the dE,N model with  $M_{\rm B}$ = $-16$\,mag
is the most extensively investigated,
the $a_{\rm dm}$ of this model is hereafter referred to as $a_{\rm dm,0}$.  
In order to clarify  the importance of the dark matter
halo profiles of dE,Ns in the galaxy threshing process, 
we also investigate models with NFW dark matter profiles
with a scale radius of $a_{\rm s}$.
For the scale radius of a cluster's dark matter halo 
to be distinguished from that of a galactic dark matter halo,
we here use $a_{\rm s}$ rather than the original definition of $r_{\rm s}$.
Since there are no results on the scale radius $a_{\rm s}$ for dark
matter halos of typical dwarfs (with $M_{\rm dm}$   $\sim$ $10^9$ 
$M_{\odot}$) in NFW, direct comparison between $a_{\rm s}$ and  $a_{\rm 
dm}$ for a given total mass of a dark matter halo is not possible.
If we assume that the central density of the dark matter halo
with $M_{\rm dm}$  =  4 $\times$ $10^9$ $M_{\odot}$
(corresponding to the above $M_{\rm B}$ = $-16$ mag dE,N model)
is the same as that of the halo with $M_{\rm dm}$  =  3.2 $\times$  
$10^{11}$ $M_{\odot}$  (corresponding to model 1 in Table of NFW),
$a_{\rm s}$ is estimated to be $\sim$ $0.7 a_{\rm dm}$.
The CDM model predicts that dark matter halos with lower masses can 
show higher central densities (NFW), and $a_{\rm s}/a_{\rm dm}$ can be 
smaller than 1 for a given dE,N dark matter halo mass. Guided by this 
simple estimation, we investigate the models with 
$a_{\rm s}$ = $a_{\rm dm,0}$, 0.5$a_{\rm dm,0}$, and 0.25$a_{\rm dm,0}$.

%\begin{equation}
%{\rho}_{\rm nfw}(r)=\frac{\rho_{nfw,0}}{(r/a_{\rm s})(1+r/a_{\rm s})^2},
%\end{equation}

The mass (luminosity) and the scale length of the stellar envelope in a dE,N
is modelled according to the observed scaling relation of Ferguson \& 
Binggeli (1994): 
\begin{equation}
{\rm log}r_{0} [{\rm pc}] = -0.2 M_{\rm B} - 0.3 
\end{equation} 
for bright dwarfs ($M_{\rm B}$ $<$ $-16$) and 
\begin{equation}
{\rm log}r_{0} [{\rm pc}] = -0.02 M_{\rm B} +2.6 
\end{equation} 
for faint dwarfs ($M_{\rm B}$ $\ge$ $-16$), where
$r_{0}$ and $M_{\rm B}$ are the scale length of the exponential
profile and the absolute $B-$band magnitude, respectively.
The projected density of the envelope with $M_{\rm B}$ and the total mass
of $M_{\rm dw}$ is represented by an exponential profile with a scale 
length $a_{\rm dw}$ that is exactly the same as $r_{0}$ in the above 
equation for the value of $M_{\rm B}$. The ratio of the total mass of the 
envelope to the total luminosity of the envelope is set to be 2.
The projected density profile of the nucleus with mass $M_{\rm n}$ 
is represented by a King model (King 1964) with a core radius of 
$a_{\rm n}$ and a central concentration parameter $c$ of 1.0. 

We estimate the velocities of both the dark matter halo particles and 
the stellar envelope particles from the gravitational potential at the 
positions where they are located. In detail, we first calculate the 
one-dimensional isotropic dispersion according to the (local) virial 
theorem:
\begin{equation}
{\sigma}^{2}(r)=-\frac{U(r)}{3},
\end{equation}
where $U(r)$ is the gravitational potential at the position $r$.
We then allocate a velocity to each collisionless particle (dark matter
halo and stellar particles) so that the distribution of velocities
of these particles have a Gaussian form with a dispersion equal
to ${\sigma}^{2}(r)$. We consider that the nucleus is strongly self-gravitating
and thus the one-dimensional isotropic dispersion at a given radius in the 
nucleus is determined solely by the gravitational potential of the nucleus  
(i.e., the outer dark matter halo and the envelope do not contribute 
significantly to the potential at the very center of the dE,N). 
This is consistent with recent simulation results (Oh \& Lin 2000), 
which suggest that the relatively low central velocity dispersion
of a dE,N can be due to the self-gravitating nucleus formed by the
merging of nuclear star clusters. The method of determining the value of 
the velocity dispersion from the gravitational potential at a given
radius for the nucleus is the same as that for the dark matter and the 
stellar envelope.

For all dE,N models, the mass-to-light-ratio in the $B-$band ($M/L_{\rm 
B}$) is set to be 10 (thus $M_{\rm dm}/M_{\rm dw}$ = 5). 
Therefore, if we choose $M_{\rm B}$ for a dE,N model, then its values of  
$M_{\rm dm}$, $a_{\rm dm}$,  and $a_{\rm dw}$ are automatically
determined from equations (3) and (4). 
The remaining parameters of the dE,N model are $a_{\rm n}/a_{\rm dw}$
and $M_{\rm n}/M_{\rm dw}$. These range from 0.02 to 0.2 for $a_{\rm 
n}/a_{\rm dw}$ and from 0.02 to 0.2 for $M_{\rm n}/M_{\rm dw}$,  
and we show the results of the models that have particularly important
implications for UCD formation via galaxy threshing. 
The adopted parameter values are all consistent with
those observed in dE,Ns 
(e.g., Binggeli \& Cameron 1991; Ferguson \& Binggeli 1994). 

The adopted value of 10 for $M/L_{\rm B}$ is consistent with observational
results of $some$ dEs with well know kinematical properties.
Mateo et al. (1991) estimated $M/L_{\rm B}$ as 12$\pm$4 for the Fornax dwarf
galaxy based on the precisely determined radial velocities of 44 individual 
stars of the dwarf.
The $M/L$ estimated from radial velocity dispersion in dwarf
spheroidal galaxies in the Local group 
is summarized in van den Bergh (2000): $6-13$ for the Leo I,
$\sim$ 23 for the Carina  dwarf,  $13\pm 6$ for the Sculptor,
$\sim$ 90 for the Draco, $\sim$ 12 for the Leo II,
and $\sim$ 50 for the Sextans.
Accordingly, the adopted value of 10 in the present simulations is reasonably consistent
with observations of the Local group dwarfs
except the Draco and the Sextans with
the unusually large  $M/L$.

\subsection{High-- and low-surface brightness spirals}

We also investigate the threshing processes for nucleated low-luminosity  spiral galaxies
($M_{\rm B}$ = $-16$ mag) dominated by dark matter halos
in order to investigate whether UCD formation processes depend on
the morphological properties of their hosts.
Such galaxies are modelled by assuming they have an exponential stellar 
disk (with a radial scale length of $a_{\rm disc}$ and a vertical scale 
length of 0.2$a_{\rm disc}$) which is embedded in a dark matter 
halo. The latter is represented by an SB profile with a core radius 
$a_{\rm dm}$, and a total mass 10 times larger than the stellar disk.
The value of $a_{\rm dm}/a_{\rm disc}$ is set to be 6.0 for  the 
models, 
which ensures that the model galaxies' 
rotation curves are of similar shape to that of the Galaxy.
In addition to the rotational velocity due to the gravitational
field of the disk and halo components, initial radial and azimuthal 
velocity dispersions are assigned to the disk component according
to the epicyclic theory of Binney \& Tremaine (1987) with a Toomre 
parameter value of $Q$ = 1.5. The vertical velocity dispersion at a given 
radius is set to be half the radial velocity dispersion at that point,
consistent with that observed in the Milky Way (e.g., Wielen 1977).

We present the results for models with $M_{\rm B}$ = $-16.0$ and
central surface brightnesses of ${\mu}_{0}=22$\,mag\,arcsec$^{-2}$ in $B-$band 
(hereafter referred to as the 
``HSB'' model), 24 (``LSBI'') and 26\,mag\,arcsec$^{-2}$ (``LSBII'').
For all spiral models with $M_{\rm B}$ = $-16.0$,
the physical parameters  of the nuclei are the same as those adopted
in the fiducial dE,N (FO1) model described later.
The adopted value of 10 for the mass-to-light-ratio is reasonable
for simulations of LSBs (e.g., Mihos et al. 1997). 
The gaseous component is not included in the HSB and LSB models,
because we consider that low density gas in these galaxies
can be quickly stripped from them because of ram pressure stripping
in cluster environments.

\subsection{Orbits}

The orbit of our model dE,N (and a dwarf spiral) is assumed to be 
influenced only by the gravitational potential resulting from the dark halo
component of the Fornax cluster. Since the adopted cluster potential
is spherically symmetric (not triaxial), the orbit of the dE,N  forms a 
rosette within a plane. This orbital plane defines the $x$-$y$ plane in all 
our models. 

The center of the cluster is always set to be ($x$,$y$,$z$) = 
(0,0,0) whereas the initial position of the dE,N is set to be ($x$,$y$,$z$) 
= ($R_{\rm ini}$, 0, 0). The initial velocity of the dE,N  
($v_{\rm x}$,$v_{\rm y}$,$v_{\rm z}$) is set to be (0, $f_{\rm v} V_{\rm 
c}$, 0), where $f_{\rm v}$ and $V_{\rm c}$ are the parameters controlling 
the orbital eccentricity (i.e, the larger $f_{\rm v}$ is, the more circular
the orbit becomes) and the circular velocity of the cluster at
$R$ = $R_{\rm ini}$, respectively.  

We investigate three representative values of $f_{\rm v}$: 0.25, 0.5, and 
1, the corresponding orbital eccentricities, $e_{\rm p}$, of which are  
0.77, 0.52, and 0.0 (circular), for the dE,N with  $R_{\rm ini} = 200$\,kpc 
in the FO model. Instead of $R_{\rm ini}$ and $f_{\rm v}$,
the orbital eccentricity ($e_{\rm p}$), the apocenter ($R_{\rm a}$),
and the pericenter ($R_{\rm p}$) of the orbit of a dE,N are used for describing
the parameter set of the model, because these quantities are more 
informative. 

The adopted value of 200 kpc for $R_{\rm a}$ ($R_{\rm ini}$) in the fiducial model
corresponds roughly to the projected distance 
of the observed most distant UCD from the center 
of the Fornax cluster (Drinkwater et al. 2000a, b). 
It takes $4-5$ Gyr for dE,Ns with $R_{\rm a}$ = 200 kpc
to be transformed into UCDs, as is described later. 
The time scale for morphological transformation
from dE,Ns into UCDs is longer for larger values of $R_{\rm a}$ (Bekki et al. 2001).
Therefore, we investigate $\sim$ 4.5 Gyr evolution of a dE,N for $all$ models
in order to draw robust conclusions as to whether the strong tidal field
of the Fornax cluster can be responsible for UCD formation. 
Since UCDs can not be disintegrated after their formation  for a reasonable
set of parameters for galactic nuclei in dE,Ns (Bekki et al. 2001),
we do not intend to investigate long-term ($\sim$ 10 Gyr) evolution
of dE,Ns.

\subsection{Choice of parameters}

$M_{\rm B}$ and $e_{\rm p}$ are the most important parameters in 
determining whether a dE,N can be morphologically transformed into a UCD by 
galaxy threshing (Bekki et al. 2001); we therefore focus in most
detail on the dependences our results have on these two parameters. 
The model that shows the typical behaviour of the galaxy threshing
process, and is thus the most important in this study, is referred to
as ``the fiducial model'' (FO1). This model has $M_{\rm B} = $-16\,mag,
$a_{\rm n}/a_{\rm dw}= 0.02$, $M_{\rm n}/M_{\rm dw} = 0.05$, an SB 
dark matter profile with $e_{\rm p} = 0.77$,  $R_{\rm a} = 200$\,kpc, and 
the FO profile. By changing cluster mass profiles, host morphological 
types, dE,N luminosities, the dark matter profiles of the dE,Ns 
(as parameterised by $a_{\rm n}/a_{\rm dw}$, $M_{\rm n}/M_{\rm dw}$,
$e_{\rm p}$, and  $R_{\rm a}$), we investigate the parameter dependences of 
the galaxy threshing processes. 

The parameter values and final morphologies for each model are summarized in 
the Table 1. In the first column, FO, VI, and LG represent the models with 
the Fornax cluster, Virgo cluster and Local Group mass profiles, respectively. 
Column 8 describes the final morphological properties of the dE,Ns and dwarf 
spirals after 4.5\,Gyr orbital evolution. Here, ``UCD'' indicates a remnant 
whose envelope has been nearly completely stripped and yet whose nucleus
remains largely unaffected (at most $\sim$ 20 \% of the mass is stripped away); 
``dE,N'' indicates the case where both the envelope and nucleus survive;  
``no remnant'' indicates the case where both components are tidally 
stripped; and ``UCD+LSB env'' indicates a remnant that is dominated by
the central compact nucleus yet has a non-negligible fraction (more than 5\% 
of the total remnant mass) of the stellar envelope surrounding the nucleus. 

Even if the envelopes of dE,Ns are nearly completely removed by
galaxy threshing, not all of the threshed remnants can be identified
as UCDs. The most extreme case is the F07 model where  both the envelope 
and the nucleus of the dE,N are completely destroyed (and hence the  
``no remnant'' classification in Table 1). In this model, 
$a_{\rm n}/a_{\rm dw}$ is rather large (0.2) and the nucleus is initially 
less compact and more susceptible to tidal destruction by a cluster tidal 
field. The dE,Ns  with larger $a_{\rm n}/a_{\rm dw}$ ($>$ 0.2) are less 
likely to be transformed into UCDs because of the lower degree of 
self-gravitation of the nuclei. Although such complete destruction of 
dE,Ns is very important in the context of the origin of the intracluster
stellar light, we do not present here a detailed description of such 
``no remnant' models.
 
Finally, all the simulations have been carried out using GRAPE boards  
(Sugimoto et al. 1990) in the GRAPE 3 systems at Tohoku University
and the GRAPE 5 systems at the National Astronomical Observatory in Japan. 
The total number of particles used for each model are 20,000 for the dark 
matter, 20,000 for the stellar envelope, and 10,000 for the nucleus.
Multiple softening lengths are used for investigating the dynamical
evolution of the three collisionless components with different size and 
mass. For each component (e.g., nucleus), the gravitational softening 
length is chosen such that the length is the same as the mean particle 
separation at the half-mass radius of the component. The time integration of 
the equation of motion is performed by using a second-order, leap-frog 
method. Most of the calculations are set to be stopped at $T$ = 4.5\,Gyr
(corresponding to 20,000 time steps), where $T$ represents the time that
has elapsed since the simulation starts. 

\begin{figure}
\psfig{file=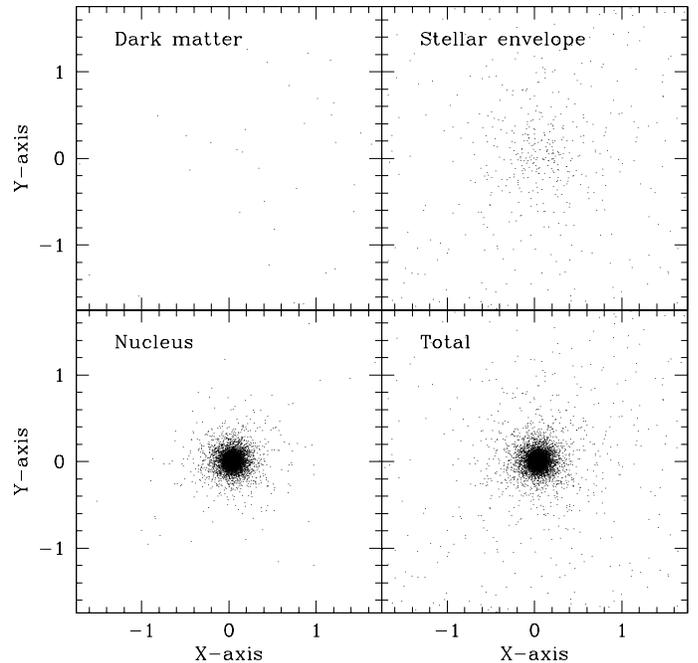,width=9.cm}
\caption{
Final mass distributions projected onto the  $x$-$y$ plane at $T$ = 4.5\,Gyr 
for the dark matter halo ({\it upper left}),
the stellar envelope ({\it upper right}), the nucleus ({\it lower left}),
and all the components ({\it lower right}).
The scale is given in units of kpc, and accordingly each frame measures 
3.5\,kpc on a side. 
}
\label{Figure 2}
\end{figure}

\section{Results}

\subsection{Fiducial model}

Figures 1 and 2 summarise the dynamical evolution of the fiducial
model and the morphological properties of the remnant in the model,
respectively. As the dE,N approaches the pericenter of its orbit,
the strong global tidal field of the Fornax cluster 
stretches the envelope of the dE'N along the direction of the dwarf's orbit 
and consequently tidally strips the stars of the envelope ($T = 1.13$\,Gyr). 
The dark matter halo, which is more widely distributed than the envelope 
due to its larger core radius, is also efficiently removed from the dE,N 
during the pericenter passage. Since the envelope (and the dark matter halo) 
loses a significant fraction of its mass during the passage of the 
pericenter, the envelope becomes more susceptible to the tidal effects of 
the Fornax cluster after the pericenter passage. Therefore, each subsequent
time the dwarf approaches the pericenter, it loses an increasingly larger
fraction of its stellar envelope through tidal stripping (compare, for
example, the $T = 2.26$ and $T = 2.83$\,Gyr time points). 
Consequently, both the envelope and the dark matter halo become
smaller,  less massive,  and more diffuse after five passages of 
the pericenter  ($T = 3.34$\,Gyr).

\begin{figure}
\psfig{file=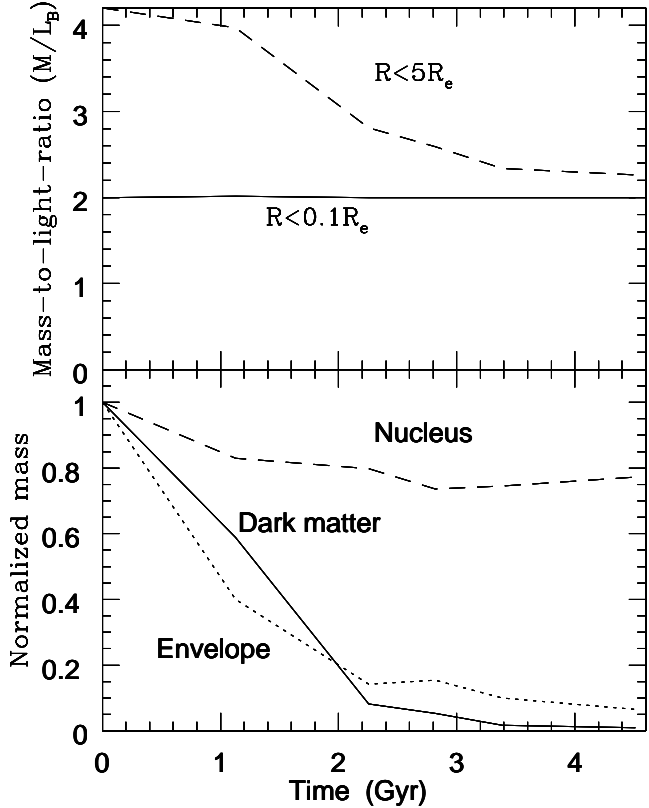,width=8.5cm}
\caption{
Time evolution of the $B$--band mass-to-light-ratio ($M/L_{\rm B}$)
({\it upper} panel) and that of the total mass normalized to the initial 
mass for each collisionless component ({\it lower} panel) in the fiducial 
model (FO1). In the {\it upper} panel, $M/L_{\rm B}$ estimated for $R$ $<$ 
$0.1R_{\rm e}$ (where $R$ and  $R_{\rm e}$ are the distance from the center 
of the dE,N and the initial effective radius of the dE,N, respectively)
and for $5R_{\rm e}$, are shown by the {\it solid} and  {\it dotted} lines,  
respectively. In the {\it lower} panel, the total mass within $R$ $<$ 
$5R_{\rm e}$, $R$ $<$ $R_{\rm e}$, and $R$ $<$ $0.1R_{\rm e}$
for the dark matter halo, the stellar envelope, and the nucleus is
shown by the {\it solid, dotted} and {\it dashed} lines, respectively.
}
\label{Figure 3}
\end{figure}

\begin{figure}
\psfig{file=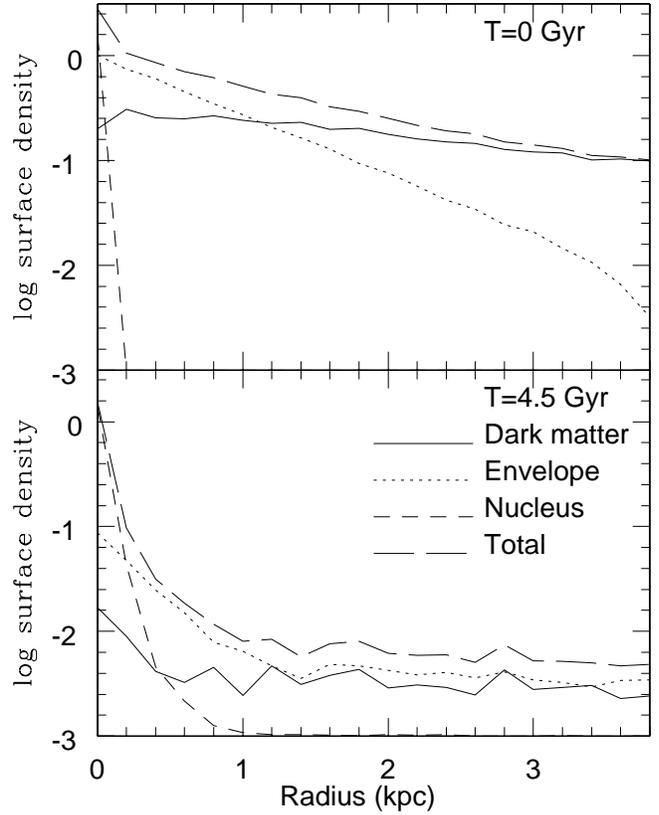,width=8.5cm}
\caption{
The projected surface density profiles for
the dark matter ({\it solid} line), the stellar envelope ({\it dotted} 
line), the nucleus ({\it short-dashed} line), and all these components 
({\it long-dashed} line) at $T = 0$\,Gyr ({\it upper} panel) and 4.5\,Gyr 
({\it lower} panel) in the fiducial model (FO1).
}
\label{Figure 4}
\end{figure}

\begin{figure}
\psfig{file=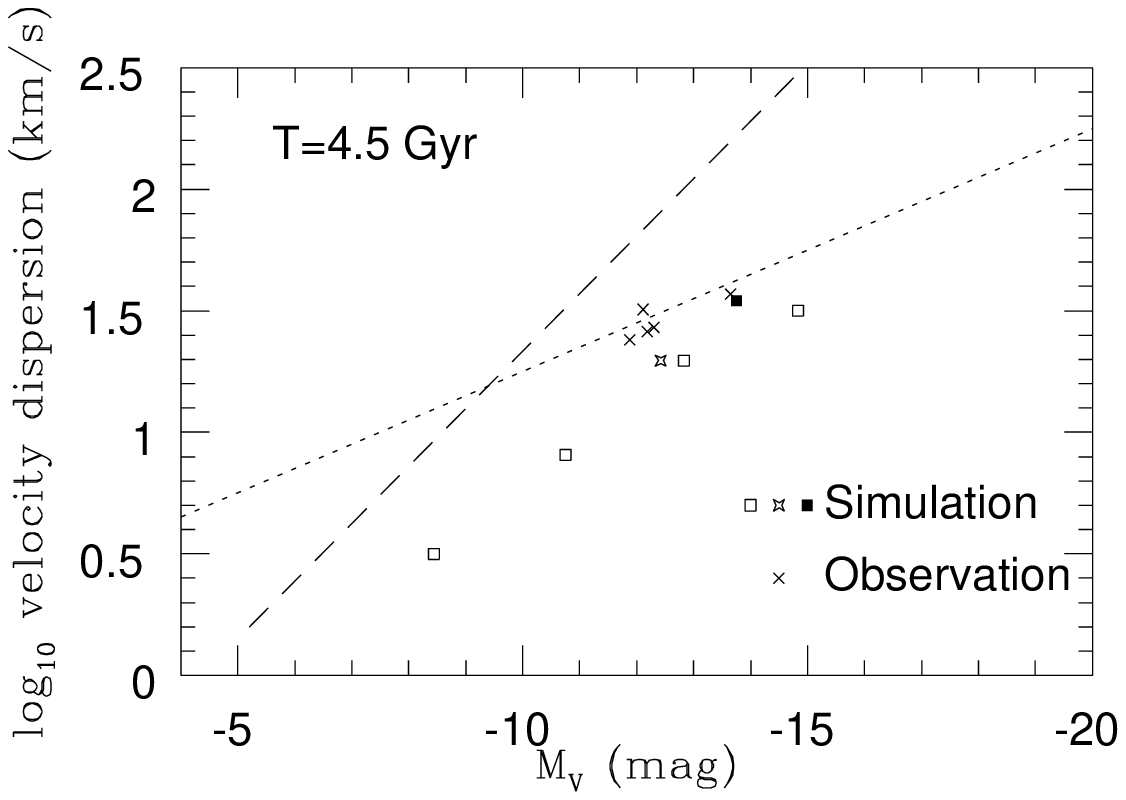,width=8.5cm}
\caption{
The distribution of the simulated UCDs on the [central velocity
dispersion, $M_{\rm V}$]--plane. Here we estimate $M_{\rm V}$ from the total 
stellar mass within $0.1R_{\rm e}$ of the UCD formed by galaxy threshing,
by assuming that the initial stellar nucleus has $M/L_{\rm B} = 2$
and $B$-$V = 0.9$ (or 0.5). 
The mass ratio of the nuclear component to the envelope one
is 0.05 for all models here.
The central velocity dispersion is expressed 
in units of km\,s$^{-1}$ but plotted on a ${\rm log}_{10}$  scale.
The simulated UCDs are represented by {\it squares} (open, filled, and 
starred); the 5 UCDs observed by D03 are plotted  
(as {\it crosses}) for comparison. The {\it open} squares represent UCDs 
formed from dE,Ns with $M_{\rm B}$ = $-12$, $-14$, $-16$, and $-18$\,mag. 
The brighter UCDs originate from brighter dE,Ns, and accordingly
 the {\it open} square at $M_{\rm V}$ = $-12.8$ mag is the UCD formed in
the fiducial model with $M_{\rm B}$ = $-16$ mag.
The {\it filled} square represents a UCD formed from a dE,N with $M_{\rm 
B} = $ $-16$ and a more massive nucleus ($M_{\rm n}/M_{\rm dw}$ = 0.2).
The {\it starred} square represents a UCD with $M_{\rm B}$ = $-16$ and 
$B-V$ = 0.5.
Long  and short dashed lines represent the GC scaling relation (e.g., Djorgovski 1993)
and Faber-Jackson one (1976), respectively. 
}
\label{Figure 5}
\end{figure}

The stripped stars form a long tidal stream along the ``rosette'' orbit 
within the orbital plane ($T = 1.13$\,Gyr). The length and shape of the 
stream depend not only on the time $T$ (the time that has elapsed since 
the simulation starts) but also on where the dE,N is located with respect 
to the center of the Fornax cluster. After a few pericenter passages,  
the stream comes to have the outer sharp edge at the apocenter of the 
dE's orbit ($T = 2.83$, 3.34, and 4.52\,Gyr). Due to the longer time 
scale for the dynamical relaxation of stars in the tidal stream,
the stars cannot be randomly dispersed into the intracluster region.
Consequently, the finer stream can still be clearly seen after 4.5\,Gyrs 
evolution of the dE,N. The developed stream is one of the observable 
predictions from the galaxy threshing model of UCD formation,
and thus the existence or non-existence of such a stream extending from
an UCD should be investigated observationally. Furthermore,
the shape of the stream might well enable us to infer the orbit of the 
dE,N.

The central nucleus, on the other hand, is just weakly influenced by the
tidal force as a result of its compact configuration. Because of its 
strongly self-gravitating nature, the nucleus loses only a small amount 
($\sim 20$\%) of its mass and thus maintains its compact morphology during 
its tidal interaction with the Fornax cluster. As a result, a very 
compact stellar system with a negligible amount of dark matter
is formed from the dE,N by $T = 4.5$\,Gyr (see Fig. 2). The total mass and 
size of the remnant are $\sim$ 3.8 $\times$ $10^6$ $M_{\odot}$ and
$\sim$ 100 pc (five times the core or scale radius), 
consistent with the observed properties of a UCD. Hence
this new study, based on fully self-consistent numerical models of dE,Ns, 
confirms the earlier results of Bekki et  al. (2001b) based on a more
simplistic model.

Figure 3 shows that the mass-to-light ratio, $M/L_{\rm B}$, decreases 
dramatically from 4.2 to 2.3  for $r < 5r_{\rm e}$. This result clearly 
explains why the UCDs are observed to have mass-to-light ratios that are
much smaller ($M/L_{\rm B} = 2$--4; D03) than those observed for dE,Ns ($\sim$ 
10) for some of the Local group dwarf (van den Bergh 2000): 
galaxy threshing is the most efficient in the outer regions of a dE,N 
where the dark matter halo dominates gravitationally. About 95\% of the 
envelope initially within $r_{\rm e}$, and 20\% of the nucleus initially 
within $0.1r_{\rm e}$ are removed from the dE,N. Figure 3 also suggests 
that a dE,N whose envelope has been partially removed (i.e., at time 
$T = 2.83$\,Gyr) has a smaller $M/L_{\rm B}$  and a larger $M_{\rm 
n}/M_{\rm dw}$ compared with the initial values. This implies that {\it if 
dE,Ns in a cluster are suffering from galaxy threshing, dE,Ns with 
smaller $M/L_{\rm B}$ are likely to show larger $M_{\rm n}/M_{\rm dw}$.}
This possible correlation may well be observed thus providing a further
test of the threshing scenario.

Figure 4 describes the final projected density profile of a UCD formed 
by galaxy threshing. Both the surface density of the dark matter and that 
of the envelope drop by more than an order of magnitude within 4.5\,Gyr.
If the initial central surface brightness of the dE,N  is 
$\mu_{B}=23$\,mag\,arcsec$^{-2}$, then the final surface brightness of the dE,N
at $5a_{\rm dw}$\, is about $\mu_{B}=29$\,mag\,arcsec$^{-2}$. Such a faint, 
low surface brightness envelope will be hard to detect, even by existing 
large ground-based telescopes. Thus the UCD formed by galaxy threshing
shows a very steep density profile with an extremely low density outer 
``halo'' ($R > 100$\,pc) dominated by the original envelope component of the
progenitor dE,N. This ``halo'' component surrounding a UCD is another 
important prediction of the threshing model. Furthermore, the total mass (or
luminosity) and the surface brightness of this faint outer ``halo'' component 
are continually reduced as the threshing process proceeds to completion.

Figure 5 shows the location of the remnant of the fiducial model on 
the [${\sigma}_{0}$, $M_{\rm V}$]--plane. 
In estimating $M_{\rm V}$ from the
total mass of the remnant and the dE,N's initial values of $M_{\rm B}$ and
$M/L_{\rm B}$, we assume that the $B$-$V$ color of the remnant is 0.9 (also
0.5) mag, which is the same as the color D03 observed for
their sample of UCDs. 
We estimate the central velocity dispersion from stars within
100 pc of the remnant of the threshed dE,N in each model.
As is shown in this Figure, the remnant of the fiducial model
has 
$M_{\rm V} = $ $-12.8$\,mag  (and $M_{\rm B} = $ $-11.9$), which is $\sim 4$\,mag  
fainter than the initial dE,N, and ${\rm log}_{10} {\sigma}_{0} = 1.3$, 
so that the location on the [${\sigma}_{0}$, $M_{\rm V}$]--plane is fairly
close to that observed for UCDs. Figure 5 also demonstrates that: (i)\,only
luminous dE,Ns ($M_{\rm B} < $ $-16$ \,mag) can become UCDs with 
$M_{\rm V} < $ $-12$ \,mag and ${\sigma}_{0} > $20\,km\,s$^{-2}$ after being
threshed, (ii)\,the remnants of dE,Ns with the more massive (or
bluer) nuclei lie closest to the location of the observed UCDs,
and (iii)\,dE,Ns which overall are more luminous, are likely to become more
luminous UCDs (if nuclear mass fraction is constant for dE,N populations) . 
The results shown in Figures 3, 4, and 5 thus confirm 
that galaxy threshing can transform {\it bright}  dE,Ns (with the initial
$M_{\rm B}$ $\sim$ $-16$ mag) into compact stellar remnants
with the structure and the kinematics similar to those observed in UCDs.

Figure 5 also implies that the UCDs observed by Drinkwater et al. (2000a, b), 
which have $M_{\rm V} < $ $-12$, must have had bright 
dE,N progenitors (with $M_{\rm B}$ $\sim$ $-16$ mag) if they were transformed into UCDs by
galaxy threshing. It also demonstrates that there can exist very compact
objects with $M_{\rm V}$ $>$ $-12$ mag 
(thus with $M_{\rm B}$ of the host dwarfs $>$ $-16$ mag) that do not lie on the GC scaling relation 
 However, in the absence of kinematical information,
such objects could easily be misidentified as GCs. Thus future central velocity 
dispersion measurements of the GC populations (identified around cluster
galaxies via recession velocity measurements alone) may well uncover such 
a population of lower luminosity UCDs, distinguishable by their displacement 
from the GC scaling relation.

\begin{figure*}
\psfig{file=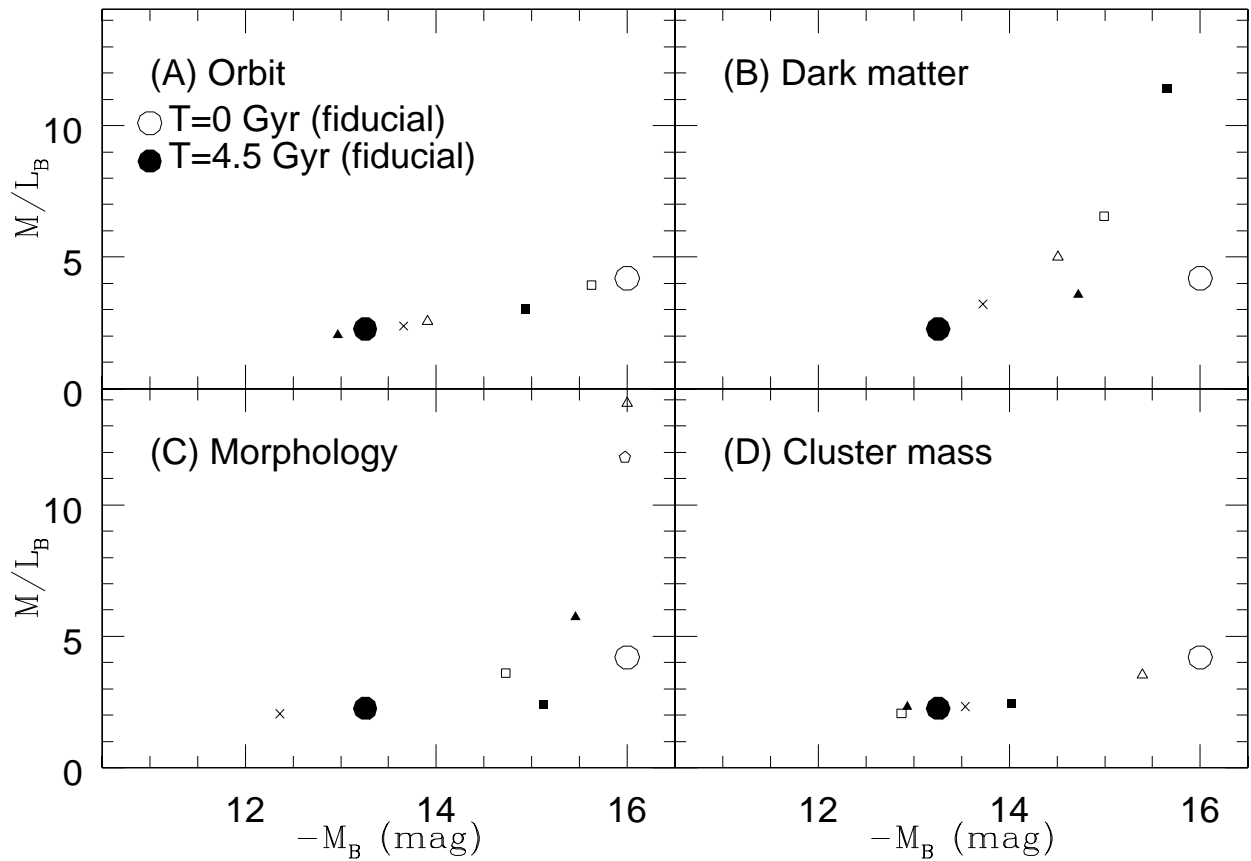,width=16.5cm}
\caption{ 
Distribution of 21 models with different parameter values on the 
[$M/L_{\rm B}$, $M_{\rm B}$]--plane. For all models, $M_{\rm B}$ and 
$M/L_{\rm B}$ are estimated for $R < 5R_{\rm e}$. In all four frames, 
the results of the fiducial model at $T = 0$\,Gyr ({\it open circle}) 
and 4.5\,Gyr ({\it filled circle}) are plotted for comparison.
{\it Upper left panel} (A): dependences on orbits for the  
FO7 (open triangles), FO8 (filled triangles), FO9 (open squares), FO10 (crosses), 
and FO11(filled squares) models. {\it Upper right panel} (B): dependences on the
dark matter profiles for the F12 (open triangles), F13 (filled triangles), F14 
(open squares), F15 (crosses), and FO16 (filled squares) models.
{\it Lower left  panel} (C): Dependences on host morphological types for the
F17 (open triangles), F18 (filled triangles), F19 (open squares), F20 (crosses),
F21 (filled squares), and F22 (open pentagons) models. 
{\it Lower right panel} (D): Dependences on cluster masses for the LG1 (open
triangles), LG2 (filled triangles), VI1 (open squares), LG3 (crosses), 
and VI2 (filled squares) models . The parameter values for each model are given
in Table 1.
The $M/L_{\rm B}$ is estimated at $T$ = 4.5 Gyr for the models. 
Note that the brighter remnants show higher $M/L_{\rm B}$. 
}
\label{Figure 6}
\end{figure*}

\subsection{Parameter dependence}

The physical properties of the threshed remnants depend strongly on
the initial parameters of the progenitor galaxies and their orbits; 
these can be seen graphically in Figure 6. 
We find the following:

(i)\,UCDs can be formed by threshing only if the orbits of the  progenitor 
dE,Ns are highly eccentric (e.g., $e_{\rm p} > $0.7 for $R_{\rm a} = 200$\,kpc 
and the FO profile). UCDs are more likely to be formed for dE,Ns orbiting 
the inner region of a cluster for a given $e_{\rm p}$. These rather 
restrictive conditions would imply that UCD formation via galaxy threshing
is expected to be a relatively rare event in clusters. 
%dE,Ns that are imperfectly being threshed show both larger  $M/L_{\rm B}$
%and smaller (brighter)  $M_{\rm B}$, which suggest that dE,Ns with smaller 
%ratios of nuclear luminosity to total one are likely to have larger 
%$M/L_{\rm B}$. 

(ii)\,If the dark matter halos of dE,Ns have the central `cuspy' cores
that recent high-resolution simulations based on CDM models predict
(e.g., NFW), these dE,Ns are less likely to be transformed into UCDs
owing to the high central dark matter density. Conversely, if the observed 
UCDs really are threshed dE,Ns, the dark matter halos of the progenitor 
dE,Ns must have a `softer' SB-like profile rather than an NFW-like profile.
Final objects of dE,Ns with more compact dark matter halos have larger 
$M/L_{\rm B}$ because of a less amount of the stripped envelopes and the 
dark matter halos. These results suggest that dE,Ns observed in the central 
regions of the present-day clusters should have more compact dark matter 
halos to survive from galaxy threshing.

(iii)\,Galaxy threshing can transform nucleated dwarf spirals into UCDs, 
provided their central surface brightness (${\mu}_{0}$) is rather low 
(${\mu}_{0} \sim 26$\,mag\,arcsec$^{-2}$). If the tidal field of a cluster 
is not strong enough to transform nucleated dwarf spirals into UCDs, it can 
transform them into nucleated S0-like objects 
due to the tidal heating of the stellar disks.  
High surface brightness nucleated spirals are very hard to be transformed 
into UCDs, even if the pericenter of the orbits is $\sim 30$\,kpc.  

(iv)\,UCD formation via dE,N galaxy threshing is more highly favoured in
more massive clusters as a result of the stronger tidal field at a given 
radius from the cluster center. For UCDs to be formed in a small group 
like the Local Group while orbiting say our Galaxy, the pericenter of the 
dE,N's orbit should be $\sim 5$\,kpc for $R_{\rm a} = 50$\,kpc (11\,kpc for 
$R_{\rm a} = 100$\,kpc). This result implies that if UCDs exist around a 
luminous galaxy (like the Galaxy and M31), the pericenter of the orbits is
likely to be within the optical radius of the galaxy ($\sim 20$\,kpc). 
This result provides some clue as to the origin of giant globular clusters 
such as $\omega$Cen and G1, which are located in the Local Group and could 
belong to the same class of compact object as UCDs.

\begin{figure}
\psfig{file=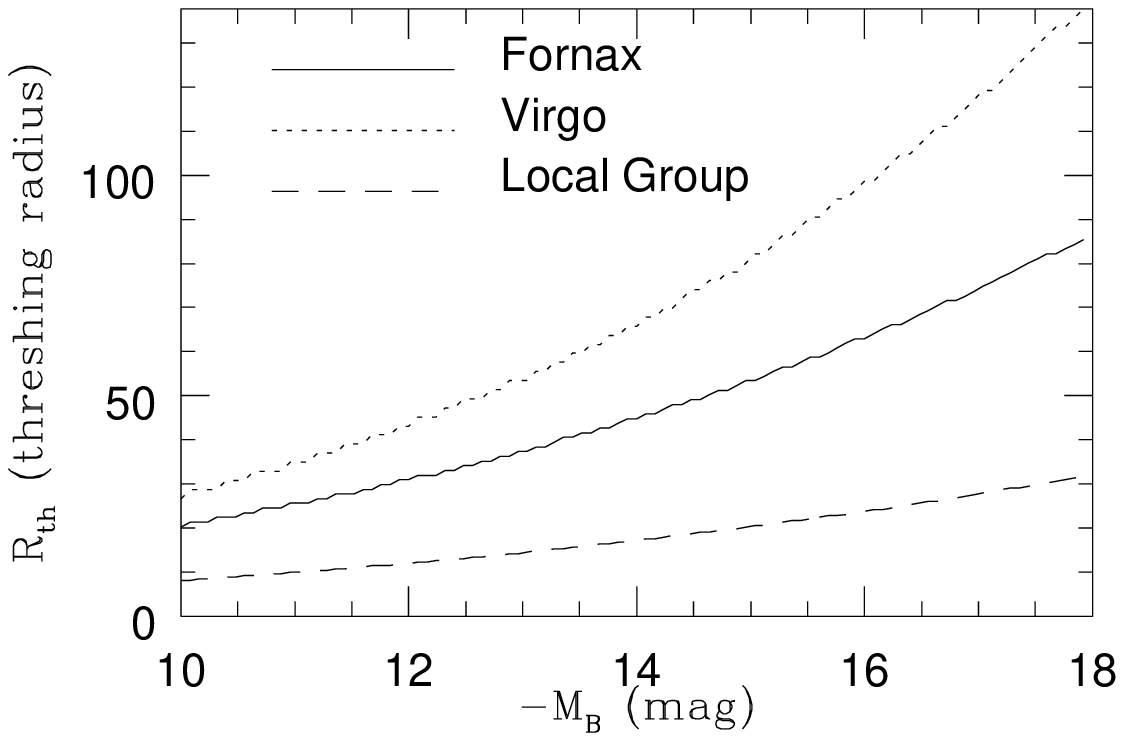,width=8.5cm}
\caption{
The dependence of the ``threshing radius'', $R_{\rm th}$ (in units of kpc), 
on $M_{\rm B}$ for our three cluster (group) models: ``Fornax'' 
({\it solid} line), ``Virgo'' ({\it dotted} line), and ``Local Group''
({\it dashed} line). A dE,N orbiting a cluster is considered to be 
transformed into a UCD (or a smaller compact stellar system) if the 
pericenter of the orbit is less than $R_{\rm th}$. An SB profile is 
adopted for the dE,N's dark matter component. The details of the 
three cluster models are given in the main text.
}
\label{Figure 7}
\end{figure}

\begin{figure}
\psfig{file=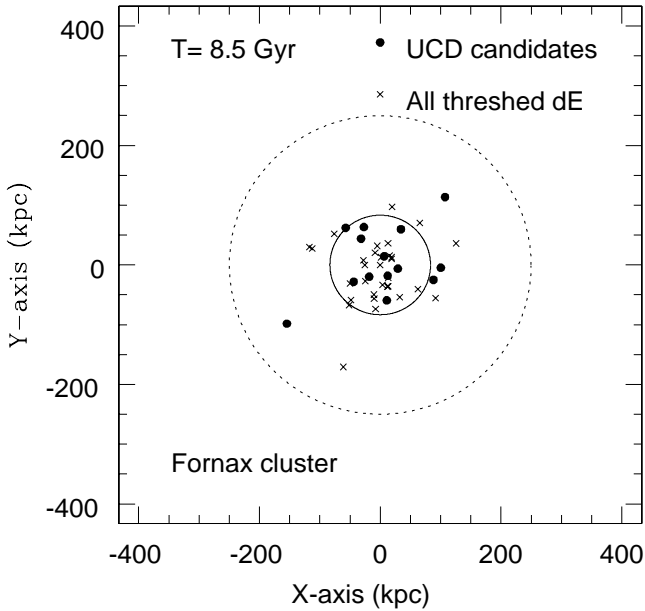,width=8.5cm}
\caption{
The projected distribution of UCDs ({\it filled} circles)
and other threshed dEs ({\it crosses}) in the Fornax model.
The {\it solid} and the {\it dotted} lines represent $r_{\rm s}$
(the cluster scale radius of the NFW profile) and $3r_{\rm s}$, 
respectively. The ``UCD candidates'' are  threshed dE,Ns with 
$M_{\rm B} < $-16\,mag, whereas the ``all threshed dE'' are 
threshed objects with $M_{\rm B} \ge $-16\,mag. This distribution is 
derived after 8.5\,Gyr of dE,N orbital evolution. 
}
\label{Figure 8}
\end{figure}

\begin{figure}
\psfig{file=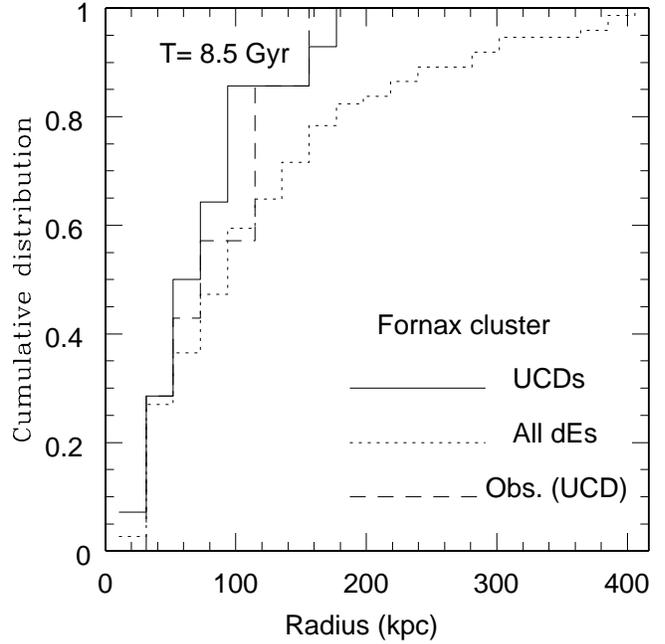,width=8.5cm}
\caption{
Cumulative radial distribution of UCDs ({\it solid} line) and all (initial) 
dwarfs with $-18 \le M_{\rm B} \le $ $-10$  ({\it dotted} line) in the Fornax 
cluster model at $T = 8.5$\,Gyr. 
For comparison, the observed distribution of UCDs (Drinkwater et al. 2000a, b)
is shown by a dashed line.
Note that UCDs are more centrally 
concentrated than dwarfs. 
}
\label{Figure 9}
\end{figure}

\section{Discussion}

\subsection{Predictions on total numbers and radial distributions
of UCDs in clusters}

Photometric and spectroscopic observations to search for UCDs in other 
nearby clusters (e.g., Virgo and Coma) are now ongoing (D03). 
The goal of these observational studies is to provide detailed 
information on (i)\,the physical properties of UCDs (e.g., total number, 
radial distribution, and luminosity function) in clusters, and 
(ii)\,the environmental dependence of these physical properties (e.g., 
Gregg et al. 2003). If these observations can be compared with the 
corresponding theoretical predictions of the galaxy threshing model,
the plausibility of the model can be checked in a more self-consistent way. 
To aid this process, we use our models to generate predictions of the 
total number of UCDs and their radial distribution as a function of 
cluster mass.

First we derive the ``threshing radius'', $R_{\rm th}$, within which
the stellar envelope of a nucleated dwarf can almost entirely be
removed by the strong tidal field of a cluster. $R_{\rm th}$ is different 
for dwarfs with different $M_{\rm B}$ in a given cluster gravitational 
potential because of the observed mass-dependent density profiles of dark 
matter halos (Salucci \& Burkert 2000). We estimate $R_{\rm th}$ for 
a dwarf orbiting a cluster by assuming that $R_{\rm th}$ is the distance
from the cluster centre at which the tidal force of the cluster is 
equivalent to the self-gravitational force of the dwarf at the core radius 
of its dark matter halo. To be more specific, $R_{\rm th}$ for a dwarf 
with $M_{\rm B}$ satisfies the following relationship:
\begin{equation} 
R_{\rm th} = 2  a_{\rm dm} {(\frac{M_{\rm cl}(R_{\rm th})}{M_{\rm dm}})}^{1/3},
\end{equation}
where $a_{\rm dm}$, $M_{\rm cl}(R_{\rm th})$, and $M_{\rm dm}$ 
are the dark matter core radius, the total mass of the cluster within 
$R_{\rm th}$, and the total mass of the dark matter in the dwarf, 
respectively. Assuming the constant $M/L_{\rm B}$ (=10) adopted in the 
present study, we can estimate $R_{\rm th}$ numerically for dwarfs with 
a given $M_{\rm B}$.

By assuming a cluster has a particular (i)\,mass-to-light-ratio ($M/L$), 
(ii)\,luminosity function, and (iii)a radial profile (for its galaxies
and dark matter halo), we can also derive the spatial distribution and 
kinematics of galaxies in the cluster. We investigate here two cluster  
models: The ``Fornax cluster model'' with total mass ($M_{\rm cl}$) of 
$7.0\times 10^{13}\,M_{\odot}$ and $M/L = 200$, and the ``Virgo cluster 
model'' with $M_{\rm cl} = 5.0\times 10^{14}\,M_{\odot}$ and  $M/L = 500$.
For comparison, we also investigate $R_{\rm th}$ for ``the Local Group model'' 
with $M_{\rm cl} = 3.0\times 10^{12}\,M_{\odot}$  and $M/L = 100$.
We adopt the Schechter function:
\begin{equation} 
{\rm \Phi(M)}=C_{\rm lf}{\Phi}^{\ast}10^{0.4(\alpha+1)(M^{\ast}-M)} 
{\rm exp}
(-10^{0.4(\alpha+1)(M^{\ast}-M)}),
\end{equation}
where ${\Phi}^{\ast}$, $\alpha$, and $M^{\ast}$ are the number density, 
faint-end slope, and characteristic absolute magnitude quantities. We
adopt the most recent observationally-determined values for these quantities: 
For $\alpha$ and $M^{\ast}$ these are -1.07 and -19.7, respectively, as
measured in the $B$-band by Efstathiou et al. (1988). 
The value of ${\Phi}^{\ast}$ is chosen according to the
values of $M_{\rm cl}$ and $M/L$ adopted for our model cluster.
For the radial galaxy number distribution of galaxies in a cluster,
we adopt a King profile with a core radius $\sim 0.6$ times
smaller than the cluster scale radius, $r_{\rm s}$, in each cluster 
model (Adami et al. 1998).
The value of $r_{\rm s}$ for each cluster
model is chosen according to the simulated values of $r_{\rm s}$ in 
Table (e.g., model 15) of NFW.

Finally, we calculate, numerically, the orbit of each galaxy (represented by a
test collisionless particle) moving within the cluster's gravitational
potential, tracking it for 8.5\,Gyr and thereby checking whether the galaxy 
comes within the threshing radius $R_{\rm th}$. In this calculation, we assume
that the mean orbital eccentricity of galaxies in a cluster is 0.6, which is
consistent with recent high-resolution cosmological simulations (Ghigna et al.
1998). 
We choose 8.5\,Gyr rather than 4.5 Gyr in order to  track evolution of dwarfs
in clusters/group, because we need to check whether or not
the dE,Ns with the apocenter distances larger than 200 kpc
(corresponding to the distance of the most distant UCD from
the center of the Fornax cluster) can be transformed into UCDs.
Since we have found that $e_{\rm p}$ should be as large as 0.7
for outer dE,Ns to be transformed into UCDs,
the predicted mean value of 0.6 in the CDM simulations (Ghigna et al. 1998)
implies that UCD formation is not so common in clusters.

The host dwarfs of UCDs should be more luminous, because the observed
mass fraction of nuclei in dE,Ns is less than 0.2 (typically 0.02)
and the UCDs are brighter
than $M_{\rm B}$ $<$ $-11$ mag.
By adopting the observationally estimated stellar masses 
(an order of $10^7$ $M_{\odot}$) by D03,
the stellar masses  of UCD progenitor dE,Ns should be
$\sim$ $10^9$ $M_{\odot}$, which corresponds to $M_{\rm B}$ = $-16$ mag
for a reasonable mass-to-light-ratio of their stellar components. 
Given the absolute magnitude range observed for UCDs ($11<-M_{\rm B}\le 13$),
we take dwarf galaxies with $16<-M_{\rm B}\le 18$ and a pericenter distance
less than $R_{\rm th}$ to be the progenitors of UCDs. We thus investigate the
total number and the radial distribution of UCDs in the two cluster models,
based on our orbital calculations for galaxies of different absolute magnitude. 
Figure 7 describes the dependence of $R_{\rm th}$ on $-M_{\rm B}$ for the two
cluster models (as well as the Local Group). Irrespective of the cluster
masses, $R_{\rm th}$ is smaller for less luminous galaxies.

Figure 8 shows the UCD distribution in the Fornax cluster model
with $M_{\rm cl} = 7.0\times 10^{13}\,M_{\odot}, M/L = 200$, and $r_{\rm
s} = 83$\,kpc. In this model, 14 UCDs can  be formed and all of them are within
3$r_{\rm s}$. This centrally concentrated UCD distribution is consistent with
that observed by Drinkwater et al. (2000a, b), although note that the observations 
do not cover the full region shown in the Figure. 
 How the UCDs are distributed spatially is further illustrated in
Figure 9 where cumulative distributions are plotted. This clearly 
demonstrates that most of the UCDs are located within 200\,kpc of the center 
of the Fornax cluster and that the UCDs are more centrally concentrated
than the dwarf galaxy population. This is qualitatively consistent with
the observations of Drinkwater et al. (2000a, b). Thus Figures 8 and  9
indicate the ``truncation radius'' within which UCDs can exist in a 
cluster. The derived number of 14 within 200 kpc is twice as much as
the observed number of UCDs within 200 kpc pf the Fornax cluster.

In Figure 10 we show the UCD distribution in the Virgo cluster model, which
has $M_{\rm cl} = 5.0\times 10^{14}\,M_{\odot}, M/L = 500$, and 
$r_{\rm s} = 226$\,kpc. Most of the 46 UCD candidates are located within
3$r_{\rm s}$, which confirms  that there is a ``truncation radius'' for 
UCDs in this cluster. As is shown in Figure 11, the cumulative distribution 
of UCDs in this model is qualitatively similar to that of the Fornax 
model. The total number of UCDs is larger and the UCDs are more widely 
distributed in the Virgo model than in the Fornax one. Thus these results 
predict that (i)\,the UCD distribution can be truncated at $R\sim 3r_{\rm 
s}$, and (ii)\,the total number of UCDs is larger in more massive clusters. 
These predictions will be readily tested by future observations.

The total number of UCDs formed by galaxy threshing in the above
simple simulation can  depend on
the total number of dE,Ns initially in a cluster 
thus on the luminosity function of galaxies
in the cluster.  Since the progenitor dwarfs  of UCDs should be more luminous
dE,Ns (typically $M_{\rm B}$ $\sim$ $-16$), 
the simulated number of UCDs depends more strongly 
on the shape of the luminosity
function around $M_{\rm B}$ $\sim$ $-16$.
Recently, Hilker et al. (2003) investigated  the luminosity function
of the Fornax cluster for fainter objects (with $M_{\rm V}$ = $-8.8$ mag)
and found the faint-end slope of the luminosity function is flat
($\alpha$ $\sim$ -1.1, See Pritchet \& van den Bergh (1999) for the luminosity
function of the Local Group of galaxies).
These recent observational results are relevant to the number of
dwarfs much fainter than UCD progenitor ones, so we think that
adopting the most recent luminosity function 
does not change the above results
so significantly.

\begin{figure}
\psfig{file=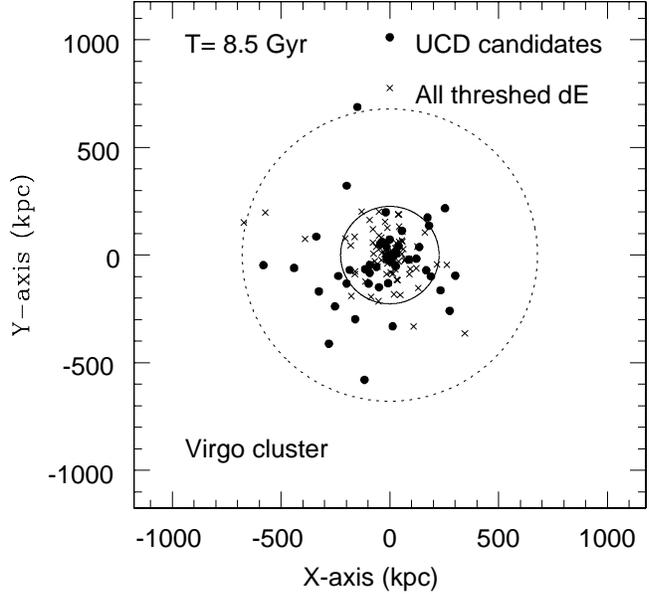,width=8.5cm}
\caption{
As for Figure 8 but for the Virgo cluster model.
}
\label{Figure 10}
\end{figure}

\begin{figure}
\psfig{file=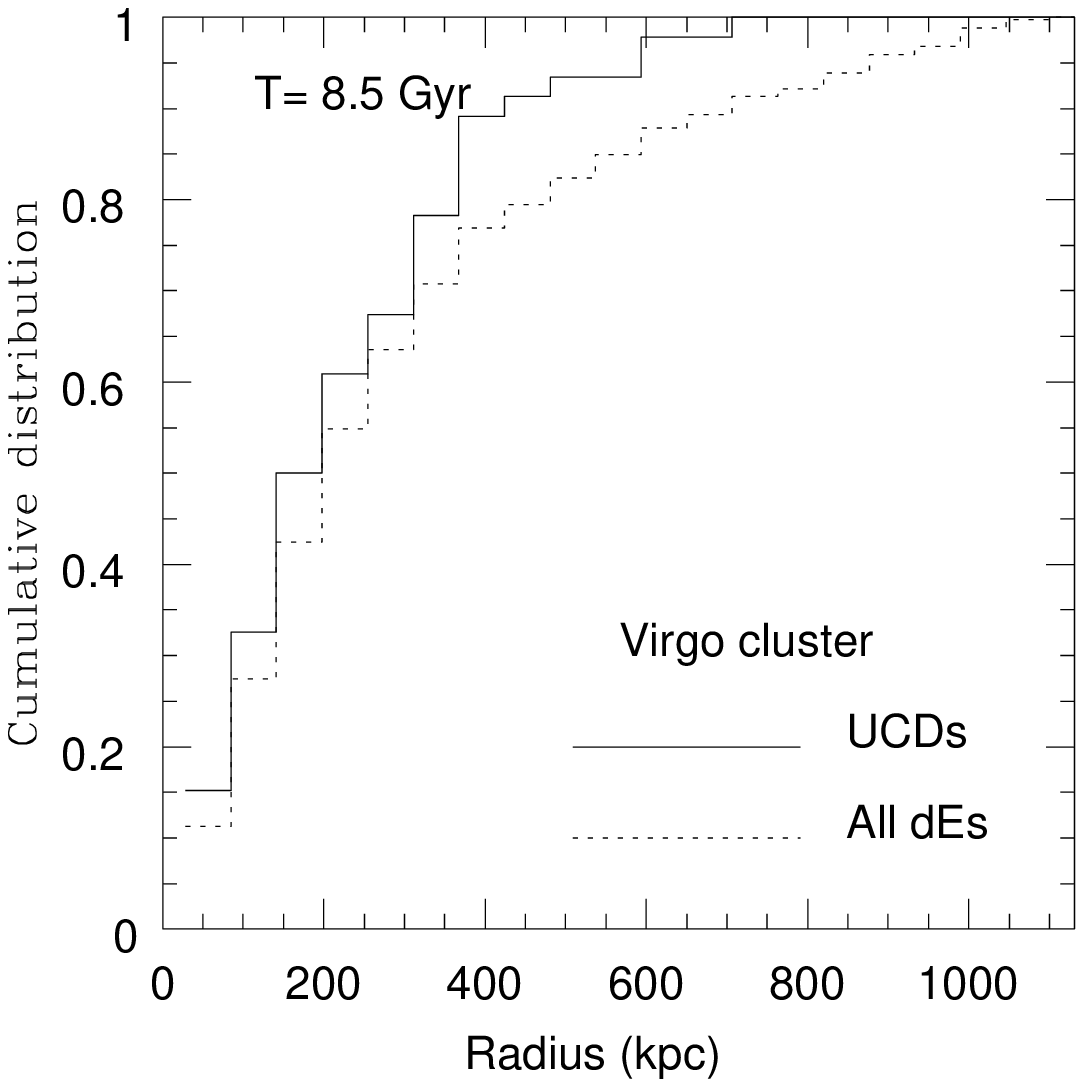,width=8.5cm}
\caption{
As for Figure 9 but for the Virgo cluster model.
}
\label{Figure 11}
\end{figure}

\begin{figure}
\psfig{file=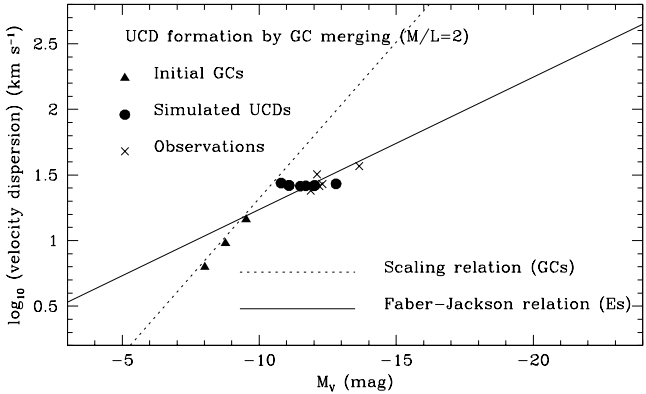,width=8.5cm}
\caption{
Distribution of UCDs formed from GC merging on the 
[central velocity dispersion, $M_{\rm V}$]--plane.
Six UCDs formed in the central regions of dEs via GC merging
are represented by the {\it filled circles}. The central velocity dispersion is
given in units of km s$^{-1}$ and plotted on a ${\rm log}_{10}$ scale. 
For comparison, the observed UCDs ({\it crosses}) and the original GCs 
({\it filled triangles}) are also plotted. The {\it solid} and the {\it dotted}
lines represent the scaling relations for GCs (Djorgovski 1993) and elliptical galaxies
(Faber \& Jackson  1976), respectively. The model and the method for
deriving $M_{\rm V}$ and the central velocity dispersion of the
UCDs formed from GC merging are given in the main text.
} 
\label{Figure 12}
\end{figure}

\subsection{Formation of stellar nuclei of dE,Ns through merging
of star clusters}

If UCDs can originate from the nuclei of nucleated dwarf galaxies, the next
question is how these nuclei are formed within the cluster environment?
Oh et al. (2000) first demonstrated that if the external tidal perturbation
of a cluster of galaxies is relatively weak for a dwarf orbiting the cluster,
dynamical friction can lead to significant orbital decay of the globular
clusters within it and the formation of compact nuclei within a Hubble timescale.
Fellhauer \& Kroupa (2002) also showed that the merging of star clusters is essential to 
the formation of compact nuclei. 
One of the key tests which can determine whether the above ``merger
scenario'' of galactic nuclei formation is plausible and realistic is to
investigate whether the scaling relations observed for UCDs (e.g., the
Fundamental Plane [${\sigma}_{0}$, $M_{\rm B}$]-- and 
[${\mu}_{0}$, $M_{\rm B}$]--relations) are consistent with those of stellar
systems formed by the merging of globular clusters. 

Bekki et al. (2002a) investigated this point by comparing the results of  
numerical simulations of GC merging in dwarfs with the recent observations 
of D03. They found there to be consistent agreement
between the scaling relations predicted for the simulated galactic nuclei 
and those observed for UCDs. Figure 12 briefly summarizes these results
and demonstrates that the galactic nuclei formed by GC merging lie off
the GC [${\sigma}_{0}$, $M_{\rm B}$]--relation and instead lie on or close
to the Faber-Jackson relation, in the same region populated by the observed
UCDs (see Bekki et al. 2002a, 2003 for details of the GC merger model).
This result implies that conversion of the orbital angular momentum
of the GCs into internal angular momentum of a single nuclear star cluster,
and energy redistribution within the merging GCs, leads to a new scaling
relation (upon which the UCDs lie) which is quite different to that defined
by the GC population. Unfortunately, the lack of any observations which 
measure velocity dispersion profiles for UCDs makes it impossible to
determine whether their kinematical properties are consistent with the
GC merger scenario. 

Clearly more extensive comparisons between the observed structural and
kinematical properties of UCDs and the simulations are needed to
determine whether their origin is in GC merging. Future high-resolution imaging
of UCDs by $HST/ACS$ will provide finer radial light distribution profiles 
and thus more precise estimates of their central surface brightness. These
can then be compared with the observed (Kormendy 1984) and simulated  
versions of the [${\mu}_{0}$, $M_{\rm B}$]--relation. Furthermore, the shapes
of the simulated nuclei (or UCDs) in the GC merger scenario depends on the
kinematics of the GC's host dwarf galaxy, such that dwarfs with more rotational
energy (i.g., larger $V/\sigma$) are  likely to have more flattened nuclei 
(Bekki et al. 2003). Therefore statistical studies of UCD shapes based on 
future high-resolution imaging will also provide a vital clue to the origin of
galactic nuclei and UCDs. 

\subsection{Origin of higher $M/L$ in UCDs}

D03's study of the structure and kinematics of the 
UCDs in Fornax has revealed that their mass-to-light ratios range from 2 to 4
in solar units, based on the assumption that they are dynamically relaxed
systems. These values are larger than the typical $M/L$ of globular
clusters ($M/L \sim 1$), and even larger than the largest globular
clusters in M31 ($M/L = 1$-2). 
The most probable reason for this would be that UCDs are embedded by
massive dark matter halos.
However, the threshing scenario predicts that since
nearly all of the dark matter can be tidally removed from dE,Ns in this 
process,  the $M/L$ of the formed UCDs should be as small as the $M/L$ of the
baryonic nuclei of the dE,Ns (Geha et al. 2002).
Therefore, if UCD are formed by galaxy threshing, the dark matter 
halos should not be responsible for the higher  $M/L$ of UCDs 
(However, we here emphasize that if UCDs are $not$ formed from galaxy threshing, 
the possibility of dark matter halos surrounding UCDs can not
be ruled out).

Accordingly, if UCDs are formed by galaxy
threshing and are predominantly baryonic, then the fact that their 
mass-to-light ratios are observed to be higher than the values typically
measured for GCs suggests one of two possibilities: One is that the stellar
populations of UCDs and GCs are quite different. The other is that
the King model mass estimator, traditionally used for GCs (Meylan et al. 2001)
but which has also been used to derive $M/L$ values for UCDs (D03), is 
inapplicable to the latter because of fundamental differences in the
structural and kinematical properties between UCDs and GCs.
   
With regard to the first possibility, we point out that the difference in  
the initial mass function (IMF) between UCDs and GCs is more likely to cause the
observed difference in $M/L$ between these two compact stellar systems 
Charlot et al. (1993)
suggested that the IMF of stellar populations formed as the result of 
a starburst can be truncated, thus resulting in a higher $M/L$. Recent
hydrodynamical simulations have demonstrated that star clusters rather than
field stars are more likely to form in the central regions of interacting and
merging  galaxies with strong starbursts because the high pressure of the
interstellar gas can trigger the collapse of giant molecular clouds 
(Bekki \& Couch 2001; Bekki et al. 2002b). Oh et al. (2000) showed that star
clusters can merge with one another to form a single nucleus in a dwarf galaxy
because of the efficient dynamical friction of the clusters. Together these
results imply that if compact nuclei are formed by the merging of young
star clusters in starbursting dwarf galaxies, then they (and hence UCD
progenitors) will have a higher $M/L$. Spectrophotometric studies which
place strong constraints on the IMF of young star clusters in starbursting
dwarf galaxies provide an important text of this first scenario.
 
With regard to the measurement of the mass-to-light ratio, we note that $M/L$ is
assumed to be proportional to the product of $r_{e}$ and ${\sigma}_{0}$ in the
formula adopted by D03. Therefore the $M/L$ of each UCD can
be overestimated if either the effective radius ($r_{e}$) or the central
velocity dispersion (${\sigma}_{0}$) is overestimated for some observational
reason. The possibility that $r_{e}$ is overestimated by a factor of 2--4 is
highly unlikely, because the observed radial profiles are fitted so well by
either a King model or de Vaucouleurs ($R^{1/4}$) profile (D03). 
On the other hand, the velocity dispersion of a UCD is very likely to
decrease with radius, like other self-gravitating systems such as elliptical
galaxies. Because D03's spectroscopic observations were made
through a slit that was large compared to the angular sizes of the UCDs, 
this would cause their central velocity dispersions to be underestimated, 
in the opposite direction to that required for an overestimation of $M/L$.
Although other systematic observational effects cannot be ruled out entirely,  
it would seem more likely that the difference between the $M/L$ for GCs and
UCDs is real and related to differences in their stellar populations.

\subsection{Physical relationship between UCDs and $\omega$Cen}

The globular cluster $\omega$Cen is the most massive such system in our Galaxy 
and is observed to have unique physical properties such as a very broad
metallicity distribution (Freeman \& Rodgers 1975) and a highly flattened
shape (Meylan 1987). These unique characteristics have been cited  
as evidence for $\omega$Cen having a very different star formation,  
chemical enrichment and structure formation history to other Galactic 
globular clusters (e.g., Majewski et al. 1999; Hilker \& Richtler 2000). 
Such highly flattened, luminous globular clusters are also known to 
reside around M31, the most notable being G1 (Meylan et al. 2001), with another 
recently discovered by Larsen  (2001).

One of the most extensively discussed formation scenarios for these giant 
GCs is that they are the surviving nuclei of ancient nucleated dwarfs
in the Local Group of galaxies. Bekki \& Freeman (2003) recently 
demonstrated that $\omega$Cen could have been formed from a nucleated 
dwarf as a result of its outer stellar envelope being completely tidally 
stripped by the Galaxy. The original nucleated dwarf would have had 
an absolute magnitude of $M_{\rm B} = $ $-14$ and been in a 
retrograde orbit with respect to the Galaxy. Although there is nearly an 
order of magnitude difference in luminosity between UCDs and giant GCs such as 
$\omega$Cen and G1, both can be  formed via the same `threshing' mechanism. 
Thus it seems plausible to propose a theoretical model in which UCDs and 
giant globular clusters belong to the same family of objects with a
common progenitor -- the nuclei of dE,Ns, -- but whose luminosities/sizes 
differ between these two classes.

There are, however, two apparent  differences between UCDs and giant GCs.
Firstly, both $\omega$Cen and G1 seem to lie on the GC [$M_{\rm B}, {\sigma}_{0}$] 
scaling relation, whereas UCDs do not. Secondly, both $\omega$Cen and G1 
are considerably flattened with flatness parameter ($\epsilon$) values 
of 0.121 (Geyer et al. 1983) and 0.2 (Meylan et al. 2001), respectively.
These values are much larger than the values typical of Galactic GCs
($\epsilon = 0.006$; White \& Shawl 1987). There is no clear evidence 
that UCDs are as flattened as $\omega$Cen and G1 (D03),
although the ability of {\it HST} imagery to precisely measure their
flatness is rather limited. This difference would imply that {\it if 
both giant GCs and UCDs originate from dE,N nuclei, the less luminous
dwarfs should have more flattened nuclei.} Future statistical studies 
(based on deep imaging at high spatial resolution) of whether there is a 
correlation between flatness and the luminosity of nuclei in dwarfs
will provide a direct test of this model.

\section{Conclusions}

We have investigated, numerically, how the global tidal field of a cluster 
transforms dE,Ns morphologically into UCDs for a range of models with 
different cluster masses and sizes. We summarise our principle results as 
follows:

(1)\,For a plausible dwarf galaxy scaling relation, the outer stellar
components of a dE,N orbiting a cluster is almost totally stripped  
as a result of the strong cluster tidal field; the nucleus, on the other
hand, manages to survive. The naked nucleus developed in this `galaxy 
threshing' process will be classified as a UCD (satisfying the same 
criteria by which UCDs were discovered in the Fornax cluster) 
providing its host was more luminous than $M_{\rm B} = $ $-16$\,mag.
The mass-to-light-ratio of the threshed dE,N decreases dramatically from 
$\sim 10$ to $\sim 2$ during the course of several passages through the 
central region of the cluster.

(2)\,For a dE,N to be transformed into a UCD by threshing, its orbit needs 
to be highly eccentric. As an example, dE,Ns with flat dark matter cores 
must have orbital eccentricities of $e_{\rm p} > 0.7$  for 
$R_{a}\sim 200$\,kpc and $e_{\rm p} > 0.4$ for $R_{a} = 100$\,kpc in the 
Fornax model. This requirement does not depend on cluster mass.

(3)\,The initial dark matter density profile of the dE,N is critical
in determining whether it is transformed into a UCD by the threshing 
process. dE,Ns embedded in dark matter halos with NFW profiles (i.e., 
`cuspy' dark matter cores) are less likely to be transformed into UCDs,
because their stellar envelopes are much more strongly bound. This implies 
that if UCDs are formed via galaxy threshing, the progenitor dE,Ns must 
have very flat, low-density dark matter cores.

(4)\,Nucleated dwarf spirals can also be transformed into UCDs, but only 
if they have a low central surface brightness 
(${\mu}_{0}\sim 26$\,mag\,arcsec$^{-2}$ in the $B$-band) and larger
orbital eccentricities ($e_{\rm p} > 0.5$). Spirals with brighter
central surface brightnesses (${\mu}_{0}\sim 24$\,mag\,arcsec$^{-2}$) can 
instead be transformed into nucleated S0s as a result of the strong 
disk heating by the cluster's tidal field. High surface brightness 
nucleated dwarf spirals with ${\mu}_{0}\sim 22$\,mag\,arcsec$^{-2}$ 
cannot be transformed into UCDs, even if they are on highly eccentric 
($e_{\rm p}\sim 0.8$) orbits.

(5)\,UCDs are more likely to be formed in more massive clusters (or 
groups), because their stronger tidal fields makes the galaxy threshing
process more efficient. The total number of UCDs formed by threshing is 
larger and the UCDs are more widely distributed in such clusters. Our
threshing model predicts that there is a ``truncation radius'' within 
which UCDs can exist in a cluster.

%\acknowledgment
\section{Acknowledgment}
We are  grateful to the referee Michael Hilker for valuable comments,
which contribute to improve the present paper.
KB,  WJC, and MJD   acknowledge the financial support of the Australian Research Council
throughout the course of this work.
All the simulations described here were performed
GRAPE 3/5 systems at  Tohoku University
and at the National Astronomical Observatory in Japan.

\begin{table*}
\centering
\begin{minipage}{185mm}
\caption{Model parameters and results of galaxy merging and interaction.}
\begin{tabular}{ccccccccc}
Model no.  
&host %
\footnote{Morphological type of the host galaxy: dE,N, Sp, HSB, and LSB represent 
nucleated dwarf ellipticals, spirals, high surface brightness galaxies, and low surface
brightness ones, respectively.}
& $M_{B}$ (mag) 
& dark matter %  
\footnote{SB and NFW represent the dark matter radial profiles by
Salucci \& Burkert (2000) and by Navarro, Frenk, \& White (1995), respectively.}
&$R_{\rm a}$ (kpc)
&$R_{\rm p}$  (kpc)
&$e_{\rm p}$ 
%&{$e_{\rm p}$% 
%\footnote{Mass ratio of newly formed stellar components (field stars and MRC) to initial gas.}}
& final morphology  % 
\footnote{UCD+LSB env. means the UCD embedded by very low surface brightness stellar 
envelope.}
& comments \\ 
FO1 & dE,N  & -16  & SB  & 200 & 27  & 0.77 & UCD & fiducial model \\
FO2 & dE,N  & -12  & SB  & 200 & 27  & 0.77 & UCD & \\
FO3 & dE,N  & -14  & SB  & 200 & 27  & 0.77 & UCD &  \\
FO4 & dE,N  & -18  & SB  & 200 & 27  & 0.77 & UCD &  \\
FO5 & dE,N  & -16  & SB  & 200 & 27  & 0.77 & UCD & 
$M_{\rm n}/M_{\rm dw}$ = 0.2\\
FO6 & dE,N  & -16  & SB  & 200 & 27  & 0.77 & no remnant & 
$a_{\rm n}/a_{\rm dw}$ = 0.2 \\
FO7 & dE,N  & -16  & SB  & 100 & 100  & 0.0 & dE,N &  \\
FO8 & dE,N  & -16  & SB  & 100 & 35  & 0.48 & UCD &  \\
FO9 & dE,N  & -16  & SB  & 200 & 200  & 0.0 & dE,N &  \\
FO10 & dE,N  & -16  & SB  & 200 & 65  & 0.52 & UCD+LSB env. &  \\
FO11 & dE,N  & -16  & SB  & 400 & 47  & 0.79 & dE,N &  \\
FO12 & dE,N  & -16  & SB  & 200 & 27  & 0.77 & dE,N &  $a_{\rm dm}$ = 0.5$a_{\rm dm,0}$ \\
FO13 & dE,N  & -16  & NFW  & 200 & 65  & 0.52 & dE,N & $a_{\rm s}$ =  $a_{\rm dm,0}$\\
FO15 & dE,N  & -16  & NFW  & 200 & 27  & 0.77 & UCD+LSB env.  & $a_{\rm s}$ =  $a_{\rm dm,0}$\\
FO14 & dE,N  & -16  & NFW  & 200 & 27  & 0.77 & dE,N  & $a_{\rm s}$ =  0.5$a_{\rm dm,0}$ \\
FO16 & dE,N  & -16  & NFW  & 200 & 27  & 0.77 & dE,N & $a_{\rm s}$ =  0.25$a_{\rm dm,0}$ \\
FO17 & Sp(HSB)  & -16  & SB  & 200 & 65  & 0.52 & Sp & ${\mu}_{0}$=22 mag arcsec$^{-2}$ \\
FO18 & Sp(LSBI)  & -16  & SB  & 200 & 65  & 0.52 & S0 & ${\mu}_{0}$=24 mag arcsec$^{-2}$ \\
FO19 & Sp(LSBI)  & -16  & SB  & 200 & 27  & 0.77 & UCD with LSB.env 
& ${\mu}_{0}$=24 mag arcsec$^{-2}$ \\
FO20 & Sp(LSBII)  & -16  & SB  & 200 & 65  & 0.52 & UCD & ${\mu}_{0}$=26 mag arcsec$^{-2}$ \\
FO21 & Sp(LSBI)  & -16  & SB  & 200 & 65  & 0.52 & S0 & 
$a_{\rm dm}/a_{\rm disc}$ = 12   
 \\
FO22 & Sp(HSB)  & -16  & SB  & 100 & 35  & 0.48 &  Sp/S0\\
VI1 & dE,N  & -16  & SB  & 200 & 27  & 0.77 & UCD &  \\
VI2 & dE,N  & -16  & SB  & 400 & 58  & 0.75 & UCD with LSB env. &  \\
LG1 & dE,N  & -16  & SB  & 200 & 21  & 0.81 & dE,N &  \\
LG2 & dE,N  & -16  & SB  & 50 & 5  & 0.81 & UCD &  \\
LG3 & dE,N  & -16  & SB  & 100 & 11  & 0.80 & UCD &  \\

\end{tabular}
\end{minipage}
\end{table*}


\begin{thebibliography}{99}



\bibitem[]{}
Adami, C., Mazure, A., Katgert, P.,  Biviano, A.
1998, A\&A, 336, 63

\bibitem[Bassino et al. 1994]{bas94}
Bassino, L. P., Muzzio, J. C., \& Rabolli, M. 1994, ApJ, 431, 634

\bibitem[]{}
Bekki, K., Couch, W. J., 2001, ApJL, 557, 19 

\bibitem[]{}
Bekki, K., Couch, W. J., Drinkwater, M. J., Gregg, M. D.
2001a, ApJL, 557, 39

\bibitem[]{}
Bekki, K., Couch, W. J., Drinkwater, M. J.
2001b, ApJL, 552, 105 

\bibitem[Bekki et al. 2002]{be02}
Bekki, K., Couch, W. J., \& Drinkwater, M. J. 2002a, in IAU 8th Asian-Pacific
regional meeting, edited by S. Ikeuchi, J. Hearnshaw, \& T. Hanawa,
(PASJ), p 245. 

\bibitem[]{}
Bekki, K., Forbes, D. A., Beasley, M. A., Couch, W. J. 2002b, MNRAS, 335, 1176

\bibitem[]{}
Bekki, K., Couch, W. J., \& Drinkwater, M. J.,  Shioya, Y. 2003, in preparation 

\bibitem[]{}
Bekki, K., Freeman, K. C. 2003, in preparation 


\bibitem[Binggeli et al. 1985]{bi85}
Binggeli, B., Sandage, A.,  Tammann, G. A. 1985, AJ, 90, 1681 

\bibitem[Binggeli \& Cameron 1991]{bi91}
Binggeli, B.,  Cameron, L. M., 1991, A\&A, 252, 27

\bibitem[Binney \& Tremaine 1987]{bi87}
Binney, J.,  Tremaine, S., 1987 in Galactic Dynamics.


%\bibitem[]{}
%Bruzual A., G.,  Charlot, S. 1993, ApJ, 405, 538

\bibitem[Burkert  1994]{bu94}
Burkert, A. 1994, MNRAS, 266, 877

\bibitem[]{}
Burkert, A. 1995, ApJL, 447, 25

%\bibitem[Carignan \& Beaulieu 1989]{bi89}
%Carignan, C.,   Beaulieu, S.  1989, ApJ, 347, 760

\bibitem[]{}
Charlot, S., Ferrari, F. Mathews, G. J., Silk, J.
1993, ApJL, 419, 57


\bibitem[]{}
Dirsch, B., Richtler, T., Geisler, D., Forte, J. C.,
Bassino, L. P., Gieren, W. P.  2003, AJ, 125, 1908

\bibitem[]{}
Djorgovski, S. 1993,
in ASP Conf. Ser. 48 The globular cluster-galaxy connection.
ed. Graeme H. Smith and Jean P. Brodie (San Francisco: ASP), p496

\bibitem[Drinkwater  et al.  2000a]{dr00a}
Drinkwater, M. J., Phillipps, S.,  Jones, J. B., Gregg, M. D.,   Deady, J. H., 
Davies, J. I., Parker, Q. A., Sadler, E. M.,  Smith, R. M. 2000a,
A\&A, 355, 900 

\bibitem[Drinkwater  et al.  2000b]{dr00b}
Drinkwater, M. J., Jones, J. B., Gregg, M. D.,  Phillipps, S.  2000b,
PASA, 17, 227 


\bibitem[]{}
Drinkwater, M. J., Gregg, M. D., Colless, M. 2001, ApJL, 548, 139

\bibitem[]{}
Drinkwater, M. J., Gregg, M. D.,  Hilker, M., Bekki, K., Couch, W. J.,
Ferguson, J. B.,  Jones, J. B.,   Phillipps, S. 2003, Nature, in press (D03)

\bibitem[]{}
Efstathiou, G., Ellis, R. S., Peterson, B. A. 1988, MNRAS, 232, 431

\bibitem[Faber 1973]{fa73}
Faber, S. M. 1973, ApJ, 179, 423

\bibitem[]{}
Faber, S. M., Jackson, R. E. 1976, ApJ, 204, 668

\bibitem[]{}
Fellhauer, M., Kroupa, P. 2002, MNRAS, 330, 642

\bibitem[Ferguson \& Binggeli  1994]{fb94}
Ferguson, H. C.,  Bingelli, B. 1994, A\&ARv, 6, 67

\bibitem[]{}
Forbes, A. F., Beasley, M. A., Bekki, K. 2002, in Sky \& Space, Aug/Sep 2002, p18

\bibitem[]{}
Freeman, K. C., Rodgers, A. W. 1975, ApJL, 201, 71


\bibitem[Freeman 1993]{fr93}
Freeman, k. C. 1993, in The globular clusters-galaxy connection,
edited by Graeme H. Smith, and Jean P. Brodie,
ASP conf. ser. 48, p608 

\bibitem[]{}
Geha, M., Guhathakurta, P., van der Marel, R. P. 2002, AJ, 125, 3073

\bibitem[]{}
Ghigna, S., Moore, B., Governato, F., Lake, G.,
Quinn, T., Stadel, J. 1998, MNRAS, 300, 146

\bibitem[]{}
Geyer, E. H.,  Nelles, B.,  Hopp, U. 1983, A\&A, 125, 359

\bibitem[]{}
Gregg, M. D., Drinkwater, M. J. 2003, in preparation

\bibitem[Harris et  al. 1995]{ha95}
Harris, W. E., Pritchet, C. J.,    McClure, R. D.,
1995, ApJ, 441, 120

\bibitem[Hilker et al.  1999]{hi95}
Hilker, M., Infante, L.,   Richtler, T. 1999, A\&AS, 138, 55

\bibitem[]{}
Hilker, M.,  Richtler, T. 2000, A\&A, 362, 895

\bibitem[]{}
Hilker, M., Mieske, S.,  Infante, L. 2003, A\&A, 397L, 9 

\bibitem[Jones et al 1997]{jo97}
Jones, C., Stern, C., Forman, W., Breen, J., David, L., Tucker, W., 
Franx, M. 1997, ApJ, 482, 143

%\bibitem[]{}
%Karick, A. M., Drinkwater, M. J.,  Gregg, M. D. 2003, submitted to MNRAS

\bibitem[King  1962]{kin62}
King, I. R. 1962, AJ, 67, 471


\bibitem[Kormendy  1977]{ko77}
Kormendy, J. 1977, ApJ, 218, 333

\bibitem[]{}
Larsen, S. S. 2001, AJ, 122, 1782

\bibitem[]{}
Lotz, J. M.; Telford, R.,
Ferguson, H. C., Miller, B. W.,
Stiavelli, M., Mack,   J. 2001, ApJ, 552, 572


\bibitem[Majewski et al.  2000]{ma00}
Majewski, S. R.  et al. 1999,
The  Galactic Halo : From Globular Cluster to Field Stars,
Proceedings of the 35th Liege International Astrophysics Colloquium,
Edited by A. Noels, P. Magain, D. Caro, E. Jehin, G. Parmentier, and A. A. Thoul.
p619

\bibitem[Mateo 1998]{ma98}
Mateo, M. 1998, ARAA, 36, 435

\bibitem[]{}
Mateo, M., Olszewski, E., Welch, D. L.,
Fischer, P., Kunkel, W. 1991, AJ, 102, 914

\bibitem[]{}
Meylan, G. 1987, A\&A, 184, 144

\bibitem[]{}
Meylan, G., Sarajedini, A., Jablonka, P., Djorgovski, S. G.,
Bridges, T., Rich, R. M. 2001, AJ, 122, 830

\bibitem[]{}
Mihos, J. C.,  McGaugh, S. S.,  de Blok, W. J. G. 1997, ApJL, 477, 79

\bibitem[Moore  1994]{mo94}
Moore, B. 1994, Nature, 370, 629


\bibitem[Navarro et al. 1996]{na96}
Navarro, J. F., Frenk, C. S.,  White, S. D. M.
1996, ApJ, 462, 563

\bibitem[Nieto \& Prugniel 1987]{ni87}
Nieto, J-.L., \& Prugniel, P. 1987, A\&A, 186, 30


\bibitem[Oh \& Lin 2000]{oh00}
Oh, K. S.,  Lin, D. N. C. 2000, ApJ, 543, 620 

\bibitem[]{}
Pritchet, C. J.,  van den Bergh, S. 1997, AJ, 118, 883

\bibitem[]{}
Salucci, P.,  Burkert, A. 2000, ApJL, 537, 9

\bibitem[Sandage et al. 1985]{sa85}
Sandage, A., Binggeli, B.,  Tammann, G. A. 1985, AJ, 90, 1759

\bibitem[Sugimoto et al. 1990]{sug90}
Sugimoto,~D., Chikada,~Y., Makino,~J., Ito,~T., Ebisuzaki,~T., 
Umemura, M. 1990, Nature, 345, 33

\bibitem[]{}
van den Bergh, S. 2000, The Galaxies of the Local Group (Cambridge University
Press).

\bibitem[]{}
White, R. E., Shawl, S. J. 1987, ApJ, 317, 246

\bibitem[Zinnecker  et al. 1988]{zi88}
Zinnecker, H., Keable, C. J., Dunlop, J. S., Cannon, R. D.,   Griffiths,  W. K.
1988, in Grindlay, J. E., Davis Philip A. G., eds, Globular cluster systems in Galaxies,
Dordrecht, Kluwer, p603




%\bibitem[]{}
%\bibitem[]{}
\end{thebibliography}
\end{document}